\documentclass[sigconf]{acmart}

\AtBeginDocument{%
  }

\usepackage[turnon]{notes}

\usepackage[turnon]{notes}

\renewcommand\footnotetextcopyrightpermission[1]{} 
\usepackage{hyperref}
\usepackage{boldline}
\usepackage{color,colortbl}
\usepackage{bigstrut}
\usepackage[ruled]{algorithm2e}
\usepackage{amsmath,amsfonts}
\usepackage{amsmath}
\usepackage{array}
\usepackage{algorithmic}
\usepackage{balance}
\usepackage{blindtext}
\usepackage{booktabs}
\usepackage{caption}
\usepackage{hyperref}
\usepackage[noabbrev]{cleveref}
\usepackage{comment}
\usepackage{enumitem}
\usepackage{eqparbox}
\usepackage{fancybox}
\usepackage{fancyvrb}
\usepackage{framed}
\usepackage{graphicx}
\usepackage{ifthen}
\usepackage{listings}
\usepackage{mathrsfs}
\usepackage{mdwmath}
\usepackage{mdwtab}
\usepackage{multirow}
\usepackage{pifont}
\usepackage{textcomp}
\usepackage{url}
\usepackage{xspace}
\usepackage[normalem]{ulem}
\usepackage[framemethod=TikZ]{mdframed}
\usepackage{caption}
\usepackage{subcaption}
\usepackage{tabularx}
\usepackage{tcolorbox}
\tcbuselibrary{listings,skins}
\usepackage{mathtools}
\usepackage{blindtext}
\usepackage{lstautogobble}
\DeclareGraphicsExtensions{.pdf,.jpeg,.png}
\graphicspath{{figures/}}

\newlength\Linewidth
\def\findlength{\setlength\Linewidth\linewidth
\addtolength\Linewidth{-4\fboxrule}
\addtolength\Linewidth{-3\fboxsep}
}
\newenvironment{examplebox}{\par\begingroup
  \setlength{\fboxsep}{3pt}\findlength
  \setbox0=\vbox\bgroup\noindent
  \hsize=0.90\linewidth
  \begin{minipage}{0.90\linewidth}\normalsize}
    {\end{minipage}\egroup
    \textcolor{gray20}{\fboxsep1.2pt\fbox
     {\fboxsep5pt\colorbox{gray05}{\normalcolor\box0}}}
    \endgroup\par\noindent
    \normalcolor\ignorespacesafterend}

\newcounter{RQACounter}

\usetikzlibrary{shadows}
\usepackage{graphics}
\newmdenv[
    tikzsetting= {fill=blueish},
    skipabove=0.33em,
    skipbelow=0.33em,
    linewidth=1pt,
    innerleftmargin=4pt,
    innerrightmargin=4pt,
    innertopmargin=2pt,
    innerbottommargin=2pt,
    linecolor=gray95,
    roundcorner=2pt, 
    shadow=true,
    shadowsize=4pt,
    shadowcolor=gray95
]{questionbox}

\newmdenv[
    tikzsetting= {fill=greenish},
    skipabove=0.33em,
    skipbelow=0.33em,
    linewidth=1pt,
    innerleftmargin=4pt,
    innerrightmargin=4pt,
    innertopmargin=2pt,
    innerbottommargin=2pt,
    linecolor=gray95,
    roundcorner=2pt, 
    shadow=true,
    shadowsize=4pt,
    shadowcolor=gray95
]{answerbox}

\newmdenv[
    skipabove=0.33em,
    skipbelow=0.33em,
    innerleftmargin=4pt,
    innerrightmargin=4pt,
    innertopmargin=2pt,
    innerbottommargin=2pt,
]{lessonbox}

\usepackage{tikz}
\newenvironment{lesson}
{
    \begin{lessonbox}
}
{\end{lessonbox}}

\newenvironment{result}
{\begin{answerbox}}
{\end{answerbox}}

\newenvironment{question}
{\begin{questionbox}}
{\end{questionbox}}

\definecolor{javared}{rgb}{0.6,0,0} 
\definecolor{javagreen}{rgb}{0.25,0.5,0.35} 
\definecolor{javapurple}{rgb}{0.5,0,0.35} 
\definecolor{javadocblue}{rgb}{0.25,0.35,0.75} 

\newcommand\blue[1]{\textcolor[rgb]{0.00,0.00,1.00}{{#1}}}

\definecolor{blueish}{RGB}{250, 250, 255}
\definecolor{greenish}{RGB}{250, 255, 250}
\definecolor{redish}{RGB}{255, 200, 200}
\definecolor{majenta}{RGB}{255, 0, 255}
\definecolor{olive}{RGB}{81, 73, 36}
\definecolor{darkgreen}{RGB}{0, 95, 0}
\definecolor{darkred}{RGB}{50, 0, 0}

\definecolor{highlight}{RGB}{175, 255, 100}
\definecolor{gray01}{gray}{.98}
\definecolor{gray05}{gray}{0.95}
\definecolor{gray08}{gray}{0.92}
\definecolor{gray10}{gray}{0.90}
\definecolor{gray12}{gray}{0.88}
\definecolor{gray15}{gray}{0.85}
\definecolor{gray18}{gray}{0.82}
\definecolor{gray20}{gray}{0.80}
\definecolor{gray25}{gray}{0.75}
\definecolor{gray30}{gray}{0.70}
\definecolor{gray35}{gray}{0.65}
\definecolor{gray40}{gray}{0.60}
\definecolor{gray45}{gray}{0.55}
\definecolor{gray50}{gray}{0.50}
\definecolor{gray55}{gray}{0.45}
\definecolor{gray60}{gray}{0.40}
\definecolor{gray65}{gray}{0.35}
\definecolor{gray70}{gray}{0.30}
\definecolor{gray75}{gray}{0.25}
\definecolor{gray80}{gray}{0.20}
\definecolor{gray85}{gray}{0.15}
\definecolor{gray90}{gray}{0.10}
\definecolor{gray95}{gray}{0.05}

\xspace%

\newtcbox{\inlinebox}[1][]{enhanced,
 box align=base,
 nobeforeafter,
 colback=blueish,
 size=small,
 left=0pt,
 right=0pt,
 boxsep=2pt,
 #1}

\newcommand{\RQ}[1]{%
    \begin{question}
    \small
        \refstepcounter{RQACounter} \label{rq-\arabic{RQACounter}}
        \noindent\textbf{{RQ\arabic{RQACounter}.~#1}}
    \end{question}
}

\newcommand{\RS}[2]{%
    \begin{result}
        \textbf{Result {\ref{rq-#1}}:~}{\emph {#2}}%
    \end{result}
}

\renewcommand{\cref}[1]{\Cref{#1}}

\newcommand{\rom}[1]{\uppercase\expandafter{\romannumeral #1\relax}}

\newcommand{\ie}{\hbox{\emph{i.e.,}}\xspace}

\newcommand{\wrt}{\hbox{\emph{w.r.t.}}\xspace}

\newcounter{hypothesis}
\lstdefinestyle{CustomJava}{
  belowcaptionskip=\baselineskip,
  breaklines=true,
  xleftmargin=\parindent,
  language=java,
  showstringspaces=false,
  basicstyle=\scriptsize\ttfamily,
  keywordstyle=\bfseries\color{javapurple},
  commentstyle=\itshape\blue,
  identifierstyle=\blue,
  belowskip=1pt,
  numbers=left,
  gobble=0
}

\lstdefinestyle{CustomJavaWoNumbers}{
  belowcaptionskip=0.5\baselineskip,
  breaklines=true,
  xleftmargin=\parindent,
  language=java,
  showstringspaces=false,
  basicstyle=\scriptsize\ttfamily,
  keywordstyle=\bfseries\color{javapurple},
  commentstyle=\itshape\blue,
  identifierstyle=\blue,
  belowskip=0.5pt,
  numbers=none,
  gobble=0
}

\lstdefinestyle{CustomCS}{
    belowcaptionskip=\baselineskip,
    breaklines=true,
    xleftmargin=\parindent,
    language={[Sharp]C},
    showstringspaces=false,
    basicstyle=\footnotesize\ttfamily,
    commentstyle=\color{olive},
    morecomment=[l]{//}, 
    morecomment=[s]{/*}{*/}, 
    showstringspaces=false,
    morekeywords={ abstract, event, new, struct, as, explicit, null, switch, base, extern, object, this, bool, false, operator, throw, break, finally, out, true, byte, fixed, override, try, case, float, params, typeof, catch, for, private, uint, char, foreach, protected, ulong, checked, goto, public, unchecked, class, if, readonly, unsafe, const, implicit, ref, ushort, continue, in, return, using, decimal, int, sbyte, virtual, default, interface, sealed, volatile, delegate, internal, short, void, do, is, sizeof, while, double, lock, stackalloc, else, long, static, enum, namespace, string, if, ifdef, ifndef, undef, endif, var},
    keywordstyle=\bfseries\color{blue},
    identifierstyle=\color{darkred},
    stringstyle=\color{darkgreen},
    belowskip=1pt,
    numbers=left,
    backgroundcolor=\color{gray01},
    gobble=0
}

\newcommand{\linecode}[1]{\lstinline{#1}~}

\newcommand{\ignore}[1]{{}}
\def\locations{L}

\newcommand{\Comment}[1]{}

{\newcommand{\nb}[2]{}}

\newcommand{\gptth}{\textsc{gpt3}\xspace}
\newcommand{\codex}{\textsc{codex}\xspace}
\newcommand{\davinciedit}{\textsc{code-davinci-edit-1}\xspace}
\newcommand{\cushman}{\textsc{code-cushman-1}\xspace}

\newcommand{\davinci}{\textsc{codex-davinci}\xspace}
\newcommand{\davinciname}{\textsc{code-davinci-002}\xspace}
\newcommand{\codet}{\textsc{CodeT5}\xspace}
\newcommand{\codetbase}{\textsc{CodeT5-base}\xspace}

\newcommand{\overwatch}{\textsc{overwatch}\xspace}
\newcommand{\cpo}{\textsc{c3po}\xspace}
\newcommand{\grace}{\textsc{Grace}\xspace}

\newcommand{\codetu}{\textsc{codeT5-u}\xspace}
\newcommand{\codetuf}{\textsc{codeT5-uf}\xspace}
\newcommand{\codetuo}{\textsc{codeT5-uo}\xspace}

\newcommand{\paraheader}[1]{{\em \textbf{#1}}:}

\lstdefinelanguage{diff}{
  basicstyle=\ttfamily\footnotesize,
  morecomment=[f][\lstbg{red!20}]-,
  morecomment=[f][\lstbg{green!20}]+,
  morecomment=[f][\textit]{@@},
  morecomment=[f][\textbf]{//},
}

\usepackage{multirow}
\usepackage[normalem]{ulem}

\begin{document}

\title{\grace: Language Models Meet Code Edits}




\author{Priyanshu Gupta}
\email{priyansgupta@microsoft.com}
\affiliation{%
    \institution{Microsoft}
    \country{India}
}
\authornote{Both authors contributed equally to this work.}
\author{Avishree Khare}
\authornotemark[1]
\authornote{Work done while at Microsoft}
\email{akhare@seas.upenn.edu}
\affiliation{%
    \institution{University of Pennsylvania}
    \country{USA}
}
\author{Yasharth Bajpai}
\email{ybajpai@microsoft.com}
\affiliation{%
    \institution{Microsoft}
    \country{India}
}
\author{Saikat Chakraborty}
\email{saikatc@microsoft.com}
\affiliation{%
    \institution{Microsoft Research}
    \country{USA}
}
\author{Sumit Gulwani}
\email{sumitg@microsoft.com}
\affiliation{%
    \institution{Microsoft}
    \country{USA}
}
\author{Aditya Kanade}
\email{kanadeaditya@microsoft.com}
\affiliation{%
    \institution{Microsoft Research}
    \country{India}
}
\author{Arjun Radhakrishna}
\email{arradha@microsoft.com}
\affiliation{%
    \institution{Microsoft}
    \country{USA}
}
\author{Gustavo Soares}
\email{gsoares@microsoft.com}
\affiliation{%
    \institution{Microsoft}
    \country{USA}
}
\author{Ashish Tiwari}
\email{astiwar@microsoft.com}
\affiliation{%
    \institution{Microsoft}
    \country{USA}
}

\renewcommand{\shortauthors}{Gupta, Khare, Bajpai, Chakraborty, Gulwani, Kanade, Radhakrishna, Soares and Tiwari}
\acmCodeLink{https://aka.ms/GrACE-Code}
\acmDataLink{}

\begin{abstract}
Developers spend a significant amount of time in editing code for a variety of reasons such as bug fixing or adding new features.
Designing effective methods to predict code edits has been an active yet challenging area of research due to the diversity of code edits and the difficulty of capturing the developer intent. 
In this work, we address these challenges by endowing pre-trained large language models (LLMs) with the knowledge of relevant prior associated edits, 
which we call the \grace(Generation conditioned on Associated Code Edits) method.  
The generative capability of the LLMs helps address the diversity in code changes and conditioning code generation on prior edits helps capture the latent developer intent. 
We evaluate two well-known LLMs, \codex and \codet, in zero-shot and fine-tuning settings respectively. 
In our experiments with two datasets, \grace
boosts the performance of the LLMs significantly, enabling them to generate  29\% and 54\% more correctly-edited code in top-1 suggestions relative to the current state-of-the-art symbolic and neural approaches, respectively. 
\end{abstract}

\ccsdesc[500]{Software and its engineering~Software evolution}
\ccsdesc[500]{Software and its engineering~Automatic programming}
\ccsdesc[300]{Computing methodologies~Artificial intelligence}
\keywords{
Code editing, Associated edits, Large language models, Pre-trained
model, Programming language processing
}
\maketitle

\section{Introduction}
\label{sec:intro}


Maintaining and modifying existing code takes up a considerable portion of a developer's time compared to writing new code~\cite{se,software_evolution}.
Due to the high cost of software maintenance~\cite{softwaremaintenance,softwarecosts}, popular Integrated Development Environments (IDEs) have tooling to support developers as they refactor code~\cite{fowler2018refactoring, eclipse, visualstudio}, fix defects, adapt code to changes in the environment, or add support for new or changed requirements~\cite{visualstudio,resharper}.
One desirable feature is {\em{code edit suggestions}} wherein the tools use the location where the developer is editing code, and the surrounding code context, to generate candidate edits to recommend~\cite{overwatch,intellicode}.

To automate code edit suggestions, researchers have proposed several approaches to learn \emph{edit patterns} from edits in source code repositories~\cite{getafix,revisar,refazer,Kim:SystematicCodechange}. 
However, these approaches suffer from two key limitations: (1)~They focus on \emph{individual edits} and learn program transformation rules for them. 
We note that edits are not performed in isolation.
Developers make changes at one location, then jump to another, and then maybe back to the first location to make further changes~\cite{nonlinear_maintenance}.  
The edits that developers make to the code at different locations may not be identical, but they are often interrelated. 
In fact, the next edit often depends on the previously performed edits \cite{overwatch}. 
Learning one-step edit patterns limits the ability of these approaches to accurately predict the most likely next edit.
(2)~The symbolic program transformation rules can only slice and dice the existing code and compose its pieces to create code -- they \emph{cannot generate new code} whose pieces do not already occur in the existing version. 
This limits the expressiveness of these approaches in terms of the types of edits that they can predict.

Unlike symbolic program transformation rules, neural language models have the capability to generate new code that does not necessarily occur in the surrounding code context.
The pre-trained large language models (LLMs) like \codex~\cite{codex} and \codet~\cite{codet5} have been shown to be highly proficient at generating code.
In fact, they are already impacting software engineering in significant ways, e.g., through popular code completion tools like GitHub Copilot~\cite{copilot}. 
However, when it comes to editing code, without the knowledge of previous edits, these models are unable to infer the developers' intent and fail to generate code that should be used to replace existing code in the next edit.
In this work, we explore ways to predict code edits using LLMs by conditioning code generation on prior, relevant edits. We call such prior edits \emph{associated edits} and this methodology as \textit{Generation conditioned on Associated Code Edits} (\grace).

In recent times, there have been a few attempts to leverage past history of code evolution to learn to edit code~\cite{overwatch, c3po, editpro}. \overwatch~\cite{overwatch} is a symbolic technique that mines "edit sequence patterns". 
Such a pattern is essentially a program transformation rule whose application is conditioned on the prior application of some other program transformation rules. 
Being a pattern-based technique, \overwatch suffers from the inability to generate new code, and it also requires significant engineering effort to build the underlying symbolic pattern-learning engine. 
\cpo~\cite{c3po} is a neural model to predict the next edit at a location, given the edits only in the spatial vicinity of that location.
While reliance on such spatially related edits shows initial promise towards automation in code editing, the hypothesis may not always hold true\textemdash developers may edit two locations simultaneously that are far away from each other spatially~\cite{nonlinear_maintenance}. 
In this work, we attempt to relax this reliance, and
do not restrict associated edits to be the ones that occur in the spatial vicinity of the location under consideration.
We show that associated edits obtained from temporal history can also be useful.
Further, the \cpo model is a custom model that generates the edited code by copy-pasting existing code fragments and is therefore unable to generate new code (similar to the symbolic techniques including \overwatch).

EditPro~\cite{editpro} is a recent neural model that aims to learn the edit process for natural language documents and code files.
It proposes a special multi-step procedure where the model first predicts token-wise edit actions (insert, delete, etc.), which are then subsequently applied to the code. 
The edit actions requiring code generation, such as insert and replace, require a separate decoding step.
EditPro experiments with single-line edits, whereas our datasets contain multi-line edits.
Instead of training a new type of model from scratch, which can be expensive and requires a significant amount of data, \grace 
allows us to repurpose the already powerful LLMs to generate edited code. 

We demonstrate the benefits of our approach in two settings: (1)~\emph{zero-shot setting} in which the LLM is used out-of-the-box without additional training but with an informative prompt about associated edits and (2)~\emph{fine-tuning setting} in which the LLM is fine-tuned on 
data annotated with associated edits. 
In both cases, our results show significant benefits of conditioning existing LLMs on associated edits without having to pay the price of designing and training specialized models from scratch.

In our experiments, we use the code-editing benchmarks from \overwatch and \cpo. 
As the baseline LLMs, we use the \davinci model in the zero-shot setting and the \codet model (220M params) in the fine-tuning setting. 
We show that 
the use of associated edits helps boost the ability of these models to predict the next edit compared to the pre-trained models used without associated edits.
In the case of \davinci, we get improvements of 17\% and 30\% (in absolute terms) for the Overwatch and C3PO datasets, and improvement of 7.45\% and 9.64\% in the case of \codet. 
We also compare \davinci with associated-edit prompting and the fine-tuned \codet model with the \overwatch and \cpo methods on the respective datasets.
All our models substantially outperform these methods on their own datasets by a significant margin.
Our best models outperform \overwatch on its dataset by 10.92\% and \cpo on its dataset by 28.63\% (absolute): this is 28.61\% and 53.82\% relative improvement respectively.
EditPro dataset and model have not been released by the authors yet, therefore, we were unable to compare against it.
Both \overwatch and \cpo construct edited code from {\em{existing}} or past code, whereas we use LLMs that are capable of generating {\em{new}} code.
We show that this makes our approach more general, and we can predict code edits that are often out-of-scope for these approaches.

In summary, we make the following contributions:
\begin{enumerate}
\item We consider a practically important software-engineering problem of predicting code edits and propose \grace, a novel method of leveraging powerful LLMs to predict code edits by conditioning them on prior edits.
\item Through experimentation on two datasets, we show that using \grace
we can substantially improve performance of LLMs in zero-shot or fine-tuning settings.
\item 
\grace is superior to the state-of-the-art symbolic or neural methods designed specifically to handle code edits.
\item We conduct experiments to thoroughly evaluate \grace
and report insights gleaned from them.
\end{enumerate}
\section{Motivating Example}
\label{sec:motivation}

\begin{figure*}
\centering
\hspace{3ex}
\begin{subfigure}[t]{0.25\textwidth}
\lstset{
basicstyle=\scriptsize\ttfamily,
  firstnumber=1,
  morecomment=[l][\color{grey}]{@@},
  morecomment=[l][\color{darkgreen}]{+\ },
  morecomment=[l][\color{red}]{-\ },
  escapechar=@,
  breaklines=true,
  numberblanklines=false,
  showlines=true
}
\begin{lstlisting}
using System;
using System.Linq;                       
using System.Text;           ... @ \setcounter{lstnumber}{249} @
@\setcounter{lstnumber}{250}@catch (Exception)                 
      ...
\end{lstlisting}
\caption{Version $v_1$}\label{fig:v_1}
\end{subfigure}
\hfill
\begin{subfigure}[t]{0.32\textwidth}
\lstset{
basicstyle=\scriptsize\ttfamily,
  firstnumber=1,
  morecomment=[l][\color{grey}]{@@},
  morecomment=[l][\color{darkgreen}]{+\ },
  morecomment=[l][\color{red}]{-\ },
  escapechar=@,
  numberblanklines=false,
  breaklines=true,
  showlines=true
}
\begin{lstlisting}
  using System;
  using System.Linq;                       
  using System.Text;                 ... @ \setcounter{lstnumber}{249} @
- catch (Exception)@ \setcounter{lstnumber}{249} @
+ catch (SerializationException)
\end{lstlisting} 
\caption{Version $v_2$ with associated edit $\delta_{1,2}$}\label{fig:v_2}
\end{subfigure}
\hfill
\begin{subfigure}[t]{0.32\textwidth}
\lstset{
  basicstyle=\scriptsize\ttfamily,
  firstnumber=1,
  morecomment=[l][\color{grey}]{@@},
  morecomment=[l][\color{darkgreen}]{+\ },
  morecomment=[l][\color{red}]{-\ },
  escapechar=@,
  numberblanklines=false,
  showlines=true
}
\begin{lstlisting}
  using System;
  using System.Linq;
+ using System.Runtime.Serialization;
  using System.Text;                 ... @ \setcounter{lstnumber}{250} @
  catch (SerializationException)               
\end{lstlisting} 
\caption{Version $v_3$ with target edit $\delta_{2,3}$}\label{fig:v_3}
\end{subfigure}
\caption{The developer performs edit $\delta_{1,2}$ to go from Version $v_1$ of their code to Version $v_2$. This edit serves as an associated edit that helps with predicting the edit $\delta_{2,3}$ needed to go from Version $v_2$ to $v_3$.}
\label{fig:ex_0}
\end{figure*}

\begin{table*}[t]
  \caption{Comparison of different approaches on the example from Figure~\ref{fig:ex_0}}
  \label{tab:me_comparison}
  \vspace{-2ex}
  \small
  \begin{tabular}{l||l|c}
    \toprule
    Technique & Prediction & Correct? \\
    \midrule
    \cpo & \textsc{No Response} (alien insertion) & \ding{55} \\
    \overwatch & \textsc{No Response} (no matching pattern) & \ding{55} \\
    \davinciedit & \textsc{Empty Response} & \ding{55} \\
    \davinci without associated edits & \linecode{using System.Net.Http;} & \ding{55} \\
    \codet without associated edits & \linecode{using System.Threading;} & \ding{55} \\
    \davinci with associated edits (Our approach) & \linecode{using System.Runtime.Serialization;} & \ding{51} \\
    \codet with associated edits (Our approach) & \linecode{using System.Runtime.Serialization;} & \ding{51} \\
    \bottomrule
  \end{tabular}
  \vspace{-2ex}
\end{table*}

In this section, 
we motivate \grace by using a concrete code development scenario.
We further discuss how this approach differs from existing approaches.

\paraheader{Illustrative example}
Consider a developer refactoring code shown in Figure~\ref{fig:v_1} as Version $v_1$. The goal of the developer is to use 
\linecode{SerializationException} provided by the 
\linecode{System.Runtime.Serialization} namespace to get to Version $v_3$ shown in Figure~\ref{fig:v_3}. Let us say that the developer first replaces \linecode{Exception} on line 250 in Version~$v_1$ with \linecode{SerializationException} to create Version~$v_2$ shown in Figure~\ref{fig:v_2}. This edit required to go from Version~$v_1$ to Version~$v_2$ is denoted as $\delta_{1,2}$. The developer's cursor then moves to Line~3 of Version $v_2$ and our goal is to predict the next edit the developer will perform to reach Version $v_3$, namely the edit $\delta_{2,3}$. 

\paraheader{Conditioning on prior edits}
The task of predicting the edit $\delta_{2, 3}$ is non-trivial. 
The code in Version $v_2$ has some useful information; for example, the code indicates that
\linecode{SerializationException} is defined on Line~250 of Version $v_2$ but the required \linecode{System.Runtime.Serialization} namespace hasn't been imported anywhere.
This signal, however, is faintly present within 250 lines of additional spatial context and the relationship between the added Exception and the required import is lost. 
This relationship is an important piece of information that is required to insert the \linecode{using} statement
on Line~3 of Version $v_3$.
{\em{Our first key observation for improving prediction of code updates is that it should be
conditioned on related edits from the past.}} In the above scenario, we want to predict
the update to Version $v_2$ by also looking at the how Version $v_2$ was created
from Version $v_1$. The edit $\delta_{1, 2}$ is an {\em{associated edit}}. In this example,
there is just one associated edit, but in general there can be multiple previous edits
picked as associated edits. 

There has been some recent work on predicting code changes conditioned on previous changes~\cite{c3po,overwatch}. We now discuss how these approaches work on the illustrative example. Table \ref{tab:me_comparison} shows the predictions of various techniques on the target.

\paraheader{\cpo}
\cpo is a path-based edit prediction method that generates an edit script to predict subsequent edits. It uses a pointer network to pick valid target edits at $v_2$ by attending to $\delta_{1, 2}$, represented as an edit path in the AST. As these target edits can only refer to nodes in the ASTs at $v_2$ and $\delta_{1, 2}$, the pointer network does not have access to the \linecode{Serialization} token needed to be inserted on Line~3. Therefore, \cpo would filter out above-mentioned example in its training and testing pipelines categorizing it as an `alien insertion'. When \cpo finetuned on the \overwatch train set is used to predict $\delta_{2, 3}$, it \emph{incorrectly} suggests picking an existing \linecode{using} statement. 

\paraheader{\overwatch}
\overwatch is a symbolic procedure that learns (abstract syntax) tree transformation rules from example edit sequences in the training data, and then makes predictions by applying those rewrite rules. 
The above-mentioned example does not match any of the $\sim50$ patterns that the authors released in \cite{overwatch}. Thus, out of the box, \overwatch would not be able to provide any suggestion because of unavailability of a matching pattern for the target edit in the example. {\em{If}} we provide enough edit sequences similar to ``$\delta_{1,2}$ followed by $\delta_{2,3}$'' as training data to \overwatch, then it might learn a few edit patterns depending on the examples it gets and the order in which they are generalized. The only two useful patterns that could be learned would be either (1) ``the substitution of \linecode{Exception} by \linecode{SerializedException} is followed by importing the \linecode{System.Runtime.Serialization} namespace'', or (2) ``the substitution of \linecode{Exception} by a \textit{placeholder Type} is followed by importing a \textit{placeholder namespace}.'' While pattern (1) would return the correct response, it is an ``overfit pattern'' that does not generalize to other changes in the substituted type.
Pattern (2) is too general and cannot generate a concrete suggestion due to the unbound placeholder.


\paraheader{Using LLMs}
The approaches discussed above cannot generate the right predictions either when the target requires a new token (\cpo) or when it cannot match an existing learned pattern (\overwatch). LLMs of Code have emerged as competitive code completion tools that offer generative capabilities. This leads to our second key observation: {\em{LLMs can handle diverse editing scenarios including those that involve generation of new tokens.}} We now discuss how these models work on the illustrative example.

\paraheader{LLMs without associated edits}
%
First, let us consider how 
a modern code completion tool (based on powerful LLMs)
will attempt to predict the new code
at Line~3 of Version $v_3$.
Code completion tools, like \davinci, look at the current snapshot of the code to make predictions.
In other words, the tool will look at Version $v_2$ to predict Version $v_3$.
When we provide code from version $v_2$ to \davinci, 
it correctly predicts that
something should be imported, but it predicts an incorrect namespace.
If we use \davinciedit, the editing variant of \davinci that allows you to provide instructions for editing, the prediction continues to remain incorrect.

\paraheader{LLMs with \grace
}
Following our two key observations, we present the edit $\delta_{1, 2}$ to 
\davinci, along with Line~3 of Version $v_2$ that needs to be updated.
Now, the model successfully predicts that the updated code would be 
Line~3 of Version $v_3$. We discuss the prompt design in detail in Section 4.2.

We found that this utility of associated edits for edit prediction also extends to other models: a base \codet model fine-tuned to predict $\delta_{2, 3}$ using $v_2$ incorrectly predicts \linecode{System.Threading} while the same model fine-tuned to additionally use $\delta_{1, 2}$ to make the prediction gets the import right. 
By building a code change prediction model over a {code generation} model, we are able to 
extend the scope of edit predictions. Moreover, we are able to also perform better than the
existing works on the subset of the benchmarks that are in their scope. We discuss our quantitative performance on these benchmarks compared to \cpo and \overwatch in Sections 5.4 and 5.5 respectively. We further present a qualitative analysis of the results in Section 5.6.


\ignore{

\todo{Discuss going from versions to edits, defining edits, associated edits}

In the above scenario, one may potentially argue that it may be possible to learn 
a symbolic program transformation rule and apply it.
It is not only quite challenging to implement learning of such program transformation
rules, but it is also not always the case that there are well-defined rules. In the next
scenario, we will see a case where the rule has to be adapted to the context and applied.
\todo{Decide later if this third scenario is better or we just nuke it.}

\begin{figure*}[t]
     \centering
         \includegraphics[width=\textwidth]{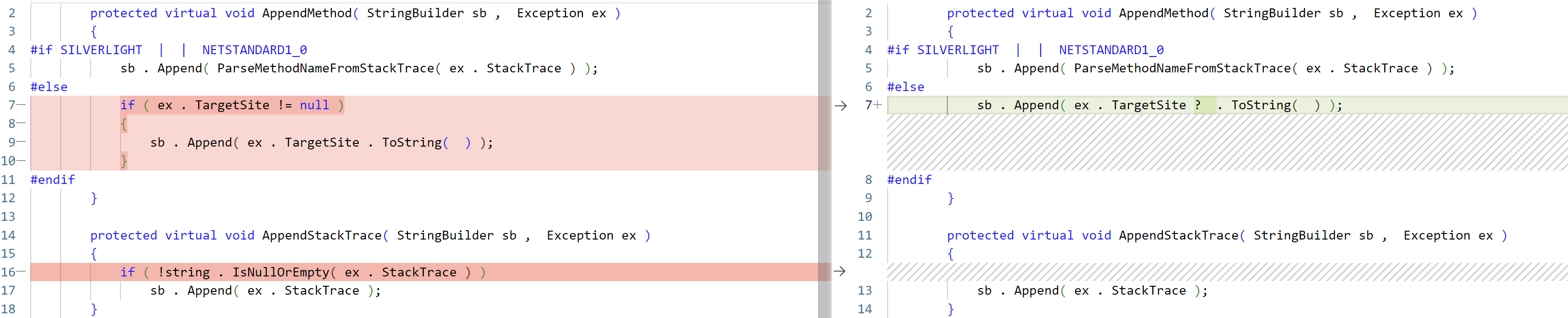}
     \caption{Predicting code change using associated edits.}
     \label{fig:motivation1}
\end{figure*}

\endignore}

\section{Associated Code Updates}
\label{sec:problem}

We define the associated code update task in this section. The associated code update task
is inspired from the EditCompletion task~\cite{c3po} and the 
edit likelihood prediction task~\cite{editpro}.

Let $v_0, \ldots, v_n$ be a sequence of versions of a source code file.
An edit $\delta_{i,j}$ is the difference between two versions, $v_i$ and  $v_j$.
We view $\delta_{i,j}$ as a function that returns $v_j$ on the input $v_i$,
i.e., $\delta_{i,j}(v_i) = v_j$. Furthermore, an {\em{associated edit}} $\Delta^i$ is $\delta_{j,k}$, for some $0 \leq j < k < n$. 
%
Given the $m$ associated edits $\Delta^1,\Delta^2,\ldots,\Delta^{m}$ and the 
version $v_{n-1}$ along with locations $\locations$ in $v_{n-1}$, the {\em{associated code update}} task is to 
predict version $v_{n}$ assuming only the locations $\locations$ in $v_{n-1}$ are updated.
We thus want to model the probability
$$
 P( v_{n} \mid \locations, v_{n-1}, \Delta^1, \Delta^2, \ldots, \Delta^m )
$$
We next make a few remarks about the problem formulation above. 
%
First, the $n-1$ versions
$v_0, v_1, \ldots, v_{n-2}$ need not necessarily match the history of the
underlying source code file.
The actual historical versions can be different, and in fact, in the 
formulation above, it is not the complete versions themselves, but the edits
$\Delta^i$ that are used in the prediction task.
The only version that is important here is the current version $v_{n-1}$.
Furthermore, the set of $m$ past edits need not even be the exhaustive set of {\em{all}} temporally consecutive edits; they could be a 
subset of the edits that have been performed so far. Hence, $\Delta^i$ doesn't necessarily have to be $\delta_{i-1, i}$.

Second, the edits are allowed to be spatially far away from each other and from the target locations $\locations$ in Version $v_{n-1}$. 
While an edit $\Delta^i$ that modifies locations close to the target locations $\locations$ is likely to be useful to include in the set $m$ of edits, edits farther away from $\locations$ may also be relevant. 
We makes no assumption on spatial locality of edits in contrast to the EditCompletion
task in~\cite{c3po}. 

\subsection{Assumptions about Sub-Problems}
\label{sec:assumptions}
Our problem formulation above abstracts away three important and challenging
related sub-problems that are crucial to build an end-to-end tool. 
These three sub-problems are:
(1)~edit localization,
(2)~edit granularity, 
and 
(3)~associated edits identification. 
 The associated code update problem formulation assumes that we have some solution for these
three related problems.

\paraheader{Edit localization}
The edit localization problem seeks to find the locations $\locations$ where
the developer should make edits. 
How we get these locations is dependent on the  application.
For example, in an IDE, cursor location is a good indicator of 
where the developer wants to make changes. 
Another option is to build a model that predicts the next edit location given 
prior associated edits. In \overwatch~\cite{overwatch}, locations were picked based on 
whether certain learned patterns matched the code at those locations. The patterns that were
matched against were selected conditioned on the past applications of associated edit patterns.


\paraheader{Edit granularity}
The edit granularity problem refers to the issue of defining what constitutes an ``edit''.
We assume that we have heuristics to define when a local code change qualifies as a
single edit. All changes between two versions that successfully parse can be used
as a definition of a single edit, as in the work~\cite{bluepencil}.
Another heuristic could be to combine all changes that occur within a small spatial vicinity
 of each other (in a commit) as a single edit~\cite{c3po}.

\paraheader{Associated edits}
The associated edits problem seeks to find edits from the past history that would be most 
useful in predicting changes at the given locations $\locations$ in the current version
$v_n$ of the source code file.
Edits that are spatially close to the target locations $\locations$ are likely relevant~\cite{c3po}.
Similarly, edits that are temporally close -- that is, edits that happened in the recent past --
are also likely candidates for being relevant. 
We can use some combination of temporal and spatial proximity to obtain a candidate
set of relevant edits~\cite{overwatch}. 
For predicting updates on a target location, the temporally-proximal edits
can indicate the developer's editing intent and the spatially-proximal edits 
can assist in providing meaning to the target snippet.
We can even further selectively choose from the edits in the spatio-temporal vicinity of
the target locations using the approach in
a recent work that mines relevant edits based on their syntactic structure and their 
likelihood of occurring together~\cite{overwatch}. We can use any or all of these
approaches to construct the set of relevant edits.
Our goal is to show that even when the relevant edits are {\em{heuristically}} generated,
using them for associated code update predictions can be very beneficial.

\subsection{Related Problem Formulations}

Existing auto-regressive LLMs, such as, \gptth and \codex, predict completions for a given prompt. 
If the prompt contains the current version $v_{n-1}$ of the artifact, then 
these LLMs predict text that is meant to be {appended} to $v_{n-1}$ to generate the new version $v_{n}$.
These models rely on the text in the spatial vicinity of the change-locations $\locations$
to make predictions. In our terminology, these models are modeling the probability 
$P(v_n \mid v_{n-1})$. This is clearly different from the problem we are considering.
We demonstrate that the associated code update formulation yields a simple yet effective way of improving 
LLM performance on software development tasks. 

The EditCompletion task in~\cite{c3po} is formalized as a study of
$P(\delta_{2,3} \mid \locations, v_2, \delta_{0,1}, \delta_{1,2})$ 
where the two given edits are edits performed
in the spatial vicinity of the current location, one before and one after the current location. 
The EditCompletion task does not consider relevant edits that may be spatially distant. 
Our problem formulation is a generalization of EditCompletion problem, and in fact, we use the
benchmarks from~\cite{c3po} for evaluation. As discussed in the introduction, our approaches are different too.

The edit likelihood prediction problem~\cite{editpro} explicitly considers the study of
$P(v_n \mid v_0,v_1,\ldots,v_{n-1})$, but it uses 
$P(\delta_{n-1,n}  \mid v_0,v_1,\ldots,v_{n-1})$ as a way to estimate the former.
This problem differs from the associated code update problem in two ways: first, it includes the sub-problem of finding the locations $\locations$ that need to be edited as part of the larger problem, and second, it considers the entire edit history as an ordered sequence (in an auto-regressive way) whereas we focus on a small set of associated edits.


\ignore{
\paragraph{Some old unpolished text.}
( Overwatch defn ): Temporal Context are changes made by the developer recently-   may help in producing suggestions. 
This temporal context can capture the editing "intent" of the developer quite well.

The spatial context, on the other hand, represents the code around a target snippet. This can be useful in providing more information about the target snippet itself. 



\endignore}

\section{Exploiting Associated Edits}
\label{sec:approach}


We propose \grace, a technique to use pre-trained language models for solving the associated code update problem.
There are two possible ways of using these models to perform
the associated code update task. 
One approach is based on using the models as a black-box, but
with carefully designed prompts (Section~\ref{sec:prompting}).
This prompting strategy works for big LLMs, such as \gptth and \codex.
The second approach is based on fine-tuning a pretrained language model,
\codet in our case, for our specific associated code update task
(Section~\ref{sec:finetuning}).


\subsection{Pre-trained Language Models}
\label{sec:ptlm}

There is now a large collection of pre-trained language models.
These models are pre-trained on data collected from millions of webpages,
and treat all data as a sequence of tokens, which is the natural
choice for representing natural language text~\cite{t5,gpt,gpt2}. 
The reason for the popularity of this class of models is that they
exhibit ability to perform multiple different tasks with just
some instructions and zero examples (zero-shot task transfer) 
even though they are not explicitly trained for these tasks. 

It was observed that pre-trained LLMs 
are not as effective when working
with code because code has strict syntactic and semantic correctness
requirements. Different representations for code and code edits
have been developed and models have been trained to work with those
representations~\cite{code2vec,c3po}. However, as the size of 
pretrained language models has grown, their zero-shot performance 
across tasks has improved. Moreover, these models have also shown the ability to 
perform a new task given just a few demonstrations 
in the prompt (few-shot learning)~\cite{gpt3}. 
Using their zero-shot and
few-shot learning capabilities, these models are now being used
successfully on tasks that involve understanding, manipulating, or
generating code~\cite{codet5,codex,copilot} while still viewing code just
as text (and not as an abstract syntax tree, for example).
%
\subsection{Prompting LLMs}
\label{sec:prompting}

We experimented with a few different prompt designs and then
fixed one for our experiments. (The results were not significantly
different for other reasonable prompt designs.) Before we describe the \grace prompt,
we first describe the completion, insertion, and editing variants
of the \codex family of models~\cite{codex,codex_insert}.

The \codex family of models is available in the ``completion'', ``insertion'' and ``editing'' variants. 
The completion model takes a {\em{prompt}}, which usually contains code before a cursor location,
and predicts the code that will follow that prompt. 
Apart from the prompt, the insertion model also takes a {\em{suffix}} prompt, which usually 
contains the part of code that should come {\em{after}} the code the model predicts. 
Thus, the insertion models perform the infilling task - predict the code that should come after the
prompt but before the suffix.
Finally, the editing variant of the codex models has two different input prompts:
an {\em{input}} that is the string that needs to be edited, and an {\em{instruction}}
that tells the model how to edit the input.

We treat the associated code update task as an infilling problem and hence use the \codex \textsc{insert} family of models for our experiments. 
The reasons for this choice are as follows:
\begin{enumerate}[leftmargin=*]
    \item The insertion model allows us to include code that is spatially after the target location in the suffix.
    \item The editing variant (\davinciedit) requires instruction on how to edit the given piece of code. 
Our experiments with providing the associated edits in this instruction prompt failed to generate good results. This is possible because the editing model is better suited only for instructions given in natural language\footnote{We do not include \davinciedit in our experiments}.
\end{enumerate}

\begin{figure}
\lstset{firstnumber=1,basicstyle=\scriptsize\ttfamily}
\begin{lstlisting}
<CurrentEdit>
    <Prefix> . . . </Prefix>
    <Before> . . . </Before>
    <After>  . . . </After>
    <Suffix> . . . </Suffix>
</CurrentEdit>
<CtxEdits>
    <Edit>
        <Prefix> . . . </Prefix>
        <Before> . . . </Before>
        <After>  . . . </After>
        <Suffix> . . . </Suffix>
    </Edit>
    <Edit>...</Edit>
    . . .
</CtxEdits>
\end{lstlisting} 
\vspace{-2ex}
\caption{\grace Prompt for the associated code update task.}
\label{fig:ft_prompt}
\end{figure}

Figure~\ref{fig:ft_prompt} shows the \grace prompt we provided \codex models for the associated code update task.
Let $v_{n-1}$ be the current version of the file,
$\locations$ be the locations where code needs to be updated, and
$\delta_{0,1}, \delta_{1,2}, \ldots,\delta_{n-2,n-1}$ be the $n-1$ associated edits.
We assume that each edit $\delta_{i-1,i}$ can be partitioned in four parts:
(1) {\linecode{<Prefix>}}, which contains the fragment of code in version $v_{i-1}$ that is untouched by the edit, but occurs before
 the edited code, 
(2) {\linecode{<Before>}}, which contains the fragment of code in version $v_{i-1}$ at locations $\locations$ that is replaced by the edit,
(3) {\linecode{<After>}}, which contains the fragment of code in version $v_{i}$ at locations $\locations$ in place of {\em{before}} in $v_{i-1}$,
(4) {\linecode{<Suffix>}}, which contains the fragment of code in version $v_{i_1}$ that is untouched by the edit, but occurs after
 the edited code.
These four parts are included in the prompt for each edit as shown in Figure~\ref{fig:ft_prompt}.
The associated edits are all included within the {\linecode{<CtxEdits>}} tag.
The edit to be predicted is included inside the {\linecode{<CurrentEdit>}} tag.

In this prompt format, the current edit is written out first followed
by the associated edits. This style ensures that if the prompt gets bigger than what can fit in the input to the
model, the tokens from the associated edits are pruned. We also experimented with variants where certain
associated edits were placed before the current edit and some after depending on where they occurred spatially.
Most such changes did not cause any significant change in our experimental observations.

The insertion model is expected to predict the string that should occur between 
\linecode{<After>} and \linecode{</After>} that occurs under \linecode{<CurrentEdit>}.
The prefix of the prompt string up until \linecode{<After>} goes in the prompt,
and the suffix of the prompt string starting from \linecode{</After>} is included in the suffix prompt
of the insertion model. 


The prompt design above is reminiscent of few-shot learning prompts where the prompt contains
a few examples of the task to be performed. Technically speaking, the above prompt is not a few-shot
prompt since we are not providing one or more examples of the ``associated code update task''. However, if
we view the associated code update problem as a {\em{means of providing few-shot examples for
the ``code update task''}}, then a natural question is whether associated edit update task can just
be viewed as a few-shot prompting for code update task. We answer this question in Section~\ref{sec:experiments}.

\begin{figure}
\vspace{-3ex}
\lstset{firstnumber=1,basicstyle=\scriptsize\ttfamily}
\begin{lstlisting}
<CurrentEdit>
        <Prefix> . . . </Prefix>
        <Before> . . . </Before>
        <After> . . . </After>
        <Suffix> . . . </Suffix>
</CurrentEdit>
\end{lstlisting} 
\vspace{-2ex}
\caption{Prompt when associated edits are not used.}
\vspace{1ex}
\label{fig:no_edit_prompt}
\end{figure}
One of the central goals of the paper is to find how using associated edits compares with
not using it when predicting code updates. To enable this comparison, we need a prompt for
the case when associated edits are unavailable.
Here, we use the prompt shown in Figure~\ref{fig:no_edit_prompt}.
Specifically, we remove the \linecode{<CtxEdits>} section in Figure~\ref{fig:ft_prompt}. Note that the current code context is still available to the model in the \linecode{<Prefix>} and \linecode{<Suffix>} tags within \linecode{<CurrentEdit>}.

\begin{table}[t]
  \caption{The models used in our experiments.} 
  \label{tab:models}
  \vspace{-2ex}
  \small
  \begin{tabular}{l||l|l}
    \toprule
    Name & Base Model & Fine-tuned on \\
    \midrule
    \davinci & \davinciname & - \\
    \codetu & \codetbase & unfiltered \cpo train\\
    \codetuf & \codetu & filtered \cpo train \\
    \codetuo & \codetu & \overwatch train \\
    \bottomrule
  \end{tabular}
\end{table}

\subsection{Fine-tuning LLMs}
\label{sec:finetuning}

We now describe how we create fine-tuned models for predicting code updates {\em{with and without}}
associated edits.
We started with the \codet-\textsc{base} model~\cite{codet5,codet5_coderl}, a pre-trained encoder-decoder
Transformer model. This base model was trained on CodeSearchNet~\cite{codesearchnet} that contains
source code in 6 common programming languages, extended with
two additional C/C\# datasets from BigQuery~\cite{bigquery1}. We further fine-tuned several variants of this model
on the task of predicting code edits (see Table~\ref{tab:models}). There are two versions of each variant -- one that is fine-tuned
using the given associated edits and one that only uses the current
version of the code. The two types of fine-tuning use the same dataset and base model weights, the only 
difference being how the data was prepared. The variants are discussed in detail in Section~\ref{sec:setup}.

We prepare data for fine-tuning by turning each training example into the \grace prompt, as shown in Figure~\ref{fig:ft_prompt}.
We adapt the \codet tokenizer by adding special tokens:
\linecode{
<Prefix>, </Prefix>, <Suffix>, </Suffix>, <CurrentEdit>, </CurrentEdit>, <CtxEdit>, </CtxEdit>, <Edit>}\newline
\linecode{</Edit>, <After>, </After>, <Before>, </Before>}.
We formulate the training as a masked span prediction task where we replace 
the contents between \linecode{<After>} and \linecode{</After>} under \linecode{<CurrentEdit>}
with a sentinel token and ask the model to predict the masked span. 

When fine-tuning \codet to predict code update {\em{without using associated edits
}}, we 
use the prompt shown in Figure~\ref{fig:no_edit_prompt}.
Again, we formulate the training as a masked span prediction task replacing
the contents between \linecode{<After>} and \linecode{</After>} under \linecode{<CurrentEdit>}
with a sentinel token and asking the model to predict the masked span. 
\ignore{
\lstset{firstnumber=1}
\begin{lstlisting}
<Prefix of final edit> 
// The following piece of code is outdated.
/* <Snippet updated by n-th edit - before snippet> */
[mask]
// Here is the new version of the code.
<Snippet updated by n-th edit - after snippet>
<Suffix of final edit>
\end{lstlisting} 
\endignore}



\subsection{Deployment}
\label{sec:deployment}
We now discuss how the sub-problems discussed in Section~\ref{sec:assumptions} can potentially be solved and integrated with our approach to create an IDE-based edit prediction tool:

\emph{\textbf{Setup}:} As discussed in Section~\ref{sec:assumptions}, the editing target could be the line corresponding to the user's cursor location. \overwatch can be used to extract temporal edits from patterns that match the target location and these edits can serve as our associated edits. 

\emph{\textbf{Worklow:}} Consider a user editing code in an IDE. The tool will get triggered on the line where the user's cursor resides and the associated edits would be retrieved using \overwatch. Our edit prediction prompt will be generated as discussed in Section~\ref{sec:prompting}. The prompt will then be sent as an input to an LLM (say, \davinci) and the predicted edit (or top-k predicted edits) will be suggested to the user. We have designed an interactive tutorial to walk readers through this workflow using the various examples discussed in Section~\ref{sec:motivation} and Section~\ref{sec:discussion} (see Section~\ref{sec:data_availability} for instructions).

\section{Experiments \& Results}
\label{sec:experiments}


\subsection{Experimental Setup}
\label{sec:setup}

We use two datasets from prior work for our experiments, 
the \cpo\ dataset~\cite{c3po} and the \overwatch\ dataset~\cite{overwatch};
see Table~\ref{tab:datasets}.

\begin{table}[t]
  \caption{The datasets used for fine-tuning and testing.} 
  \label{tab:datasets}
  \vspace{-2ex}
  \small
  \begin{tabular}{l||r|r|r}
    \toprule
    Dataset & \#training & \#eval &\#test \\
    \midrule
    \cpo filtered & 39.5K& 4.4K & 5.9K \\
    \cpo unfiltered & 1.67M& 180K & 210K \\
    \overwatch & 9K & 1k &1K \\ 
    \bottomrule
  \end{tabular}
\end{table}

\paraheader{\cpo dataset}
The \cpo dataset~\cite{c3po} was created by scraping all {\em{commits}}
in 53 most popular C\# GitHub repositories. Each edit in a commit would create a 
single example, and the edits, if any, on the 10 lines above and 10 lines below
the edit would make up the associated edits. The task is to predict the code
after an edit is performed, given the code before the edit and the associated edits.
Thus, the \cpo dataset is an instance of the associated code update task,
where \emph{spatial locality} is used to define associations between edits. 
Note that the \cpo paper refers to these edits as {\em{contextual edits}} which translate to {\em{edits with spatial associations}} in our work.

The \cpo dataset was further filtered by its creators into a 
{\em{filtered \cpo dataset}} by removing ``simple'' benchmarks (e.g. those containing
only deletion or renaming).
Further, they removed all benchmarks where the target edit involved insertion of \emph{new code} as their approach cannot handle those. The filtered set was further partitioned into train, validation, and test benchmarks, containing respectively 39.5K, 4.4K, and 5.9K benchmarks; see Table~\ref{tab:datasets}. 
We used the same partitions in our evaluation.

\paraheader{\overwatch dataset}
The dataset described in~\cite{overwatch} was gathered from versions of source code files taken as they were being edited in an IDE session over two separate periods. 
In the first period, 134.5K versions were collected over 682 sessions.
In the second period, 201.1K versions were collected over 399 sessions.
The versions in the first period were mined in~\cite{overwatch} to get a set of \emph{9.9K} edit sequences which are further used to learn a collection of symbolic rules representing commonly occurring Edit Sequence Patterns. 
These learned rules are used to generate code suggestions in the second period, and they are found to be capable of producing suggestions at \emph{1048} file versions. 
For the purpose of this work, we are considering 90\% of the \emph{9.9K} edit sequences from the first period as the \emph{\overwatch training} set, keeping other 10\% as the \emph{\overwatch evaluation} set; and the points of applications as the \emph{\overwatch test} set. We discuss these datasets further in Section \ref{subseq:overwatch}.







\paraheader{Models}
We used two models as starting points. 
The first is \davinciname (referred to as \davinci in this text), a decoder-only transformer model which is a part of the \texttt{OpenAI} \textsc{gpt-3.5} series \cite{codex, codex_insert}. 
This model is presented with the prompts based on either Figure~\ref{fig:ft_prompt} or
Figure~\ref{fig:no_edit_prompt} depending if we want to use use associated edits
for edit prediction. The second model is \codet \cite{codet5, codet5_coderl}, an encoder-decoder model introduced by Salesforce in the \textsc{base} and \textsc{large} Variants. We use the \codetbase variant which has 220M parameters with 12 transformer blocks in the encoder and decoder each. This model is fine-tuned on the unfiltered \cpo training dataset to create the model
\codetu. The model \codetu is further fine-tuned on the \cpo filtered train set and \overwatch train set to create \codetuf and \codetuo, respectively; see Table~\ref{tab:models}. 
We used this two step fine-tuning process since the \overwatch training data was limited.
The fine-tuning and inference setups for these models are described below.


\paraheader{Setup for \davinci experiments}
We used the \texttt{OpenAI} public API to perform the inference experiments with the \davinci model. The {\texttt{insert}} mode of the model was used and the input was divided into {\em{prompt}} and {\em{suffix}} following our prompting strategy discussed in Section~\ref{sec:prompting}. Temperature sampling was used to generate n=5 predictions, and the temperature was set to 0.1 after evaluating multiple candidate values. The maximum length (maximum number of tokens to generate) was set to 256, stop token to \linecode{</After>} and default values were used for all other parameters.

\paraheader{Setup for \codet experiments}
For our fine-tuning experiments, we use a virtual machine with 16 AMD MI200 GPUs (each with 64GiB of vRAM), 92 CPU cores and 1594 GB of RAM. We set the input token length to 1024 tokens and truncate any longer inputs from the end. There are two steps in our fine-tuning process: fine-tuning on the unfiltered \cpo dataset followed by dataset-specific fine-tuning on the \overwatch training and \cpo filtered datasets.
For the initial fine-tuning with the unfiltered \cpo dataset, we initialize the model with the publicly released \codet-\textsc{base} weights and train it for 8 epochs with a batch size of 8 per device. The optimization is done using the Adafactor\cite{shazeer2018adafactor} optimizer with learning rate initially set to $3e^{-4}$ and gradually updated using a linear scheduler after a warmup of 500 steps. The best model weights are determined using the perplexity score by evaluating on the \cpo validation dataset at every 1000 steps. For further fine-tuning on the \overwatch training and \cpo filtered datasets, we set the initial learning rate to $1e^{-4}$, the number of warmup steps to 50 and train the model for 10 epochs while evaluating it every 50 steps. During inference, we use beam search  with a beam width of 5. 
\paraheader{Metric}
In order to stay consistent with the metrics used by papers that curated the target datasets (namely the \cpo and \overwatch datasets), we define a metric called the {\em{exact match}}.
In the experiments with the \cpo dataset, a prediction is said to be an {\em{exact match}} if it syntactically matches the ground truth modulo whitespaces.
We use {\em{Exact Match}} to also denote the percentage of cases where a prediction was an exact match. 
More details on the \overwatch dataset evaluation can be found in Section \ref{subseq:overwatch}.
In all our results, we report {\em{Exact Match}} for Top-1 predictions.




\vspace{-2.5ex}
\subsection{\grace Improves Prediction}
\label{subsec:rq1}

A key question we set out to answer was whether 
 associated edits 
help predict
future code changes. In other words:
\RQ{Does availability of associated edits improve 
code update predictions? Does the answer depend on the prediction approach?}
To answer this question, 
we tested both \davinci and \codet on both the \cpo and \overwatch test sets,
once with associated edits in the prompt and once without them.

\begin{table}[t]
  \caption{
  Associated edits
  improve code prediction.}
  \label{tab:rq1}
  \vspace{-2ex}
  \small
  \setlength{\tabcolsep}{0.2em}
  \begin{tabular}{@{\extracolsep{4pt}}l||cccc}
    \toprule
    & \multicolumn{2}{c}{\textbf{\cpo test set}} & \multicolumn{2}{c}{\textbf{\overwatch test set}} \\
\cline{2-3}
\cline{4-5}
    Model &
    \tabular{c}Without\\ assoc. edits \\ ~ \endtabular & \tabular{c}With\\ assoc. edits \\ (\grace) \endtabular &
    \tabular{c}Without\\ assoc. edits \\ ~ \endtabular & \tabular{c}With\\ assoc. edits \\ (\grace) \endtabular
    \\
    \midrule
    \davinci & 37.09 & {\bf{67.92}} & 31.81 & {\bf{49.09}}\\
    \codetu & 64.52 & {\bf{74.16}}  & 22.25 & {\bf{34.00}}\\
    \codetuf & 73.46 & {\bf{81.83}} & 40.78 & {\bf{48.23}}\\
    \bottomrule
  \end{tabular}
\end{table}





\paraheader{Results}
Table~\ref{tab:rq1} shows the Exact Match obtained when we use the different models on the different
datasets with and without associated edits.
We see that \davinci shows a 30\% absolute increase in Exact Match 
when provided associated edits than when not on the \cpo dataset, and about 17\% absolute increase
on the \overwatch dataset.
The fine-tuned \codet models showed about a 10\% absolute increase in Exact Match on both datasets.
Finally, although Table~\ref{tab:rq1} reports the trend for 2 models and one prompting style, we tried other models (including other \texttt{OpenAI} models from  \gptth and \textsc{gpt-3.5} series)
and different styling of the prompts
(for example, using C\# comments, rather than tags, 
to delineate the ``before'' and ``after'' versions),
and in every case, there was at least a 10\% absolute increase in Exact Match -- often it was much 
higher.
%

\RS{1}{Conditioning code prediction on associated edits helps, across 
models and test datasets.}
\vspace{-2ex}
\subsection{Relevance of Edits Matters}

The associated code update problem conditions code prediction on some {\em{associated}} edits.
We have informally mentioned that the associated edits should be picked based on their relevance
to the code that is being updated. Our next research question is concerned with how relevance
impacts prediction.

To motivate this research question, we first make the connection to ``few-shot prompting''.
Consider just the {\em{code update task}} -- 
predict the new version of the code given its old version. 
The difference between the ``code update task'' and ``associated code update task'' are the
associated edits. 
Now, a prompt containing an instance of the ``associated code update'' task 
begins to look a lot similar to a {\em{few-shot prompt for a code update task}}
where the {\em{associated edits}} serve the purpose of few-shot examples of code update.

It may be tempting to say that the ``associated code update'' task just combines 
some few-shot examples with a code update task. However, this view is not beneficial since
associated edits are more than just {\em{any examples of code updates}}. As discussed in Section~\ref{sec:motivation}, the associated edits contain
crucial information for performing the given code update. To validate that associated
edits are more than just code update examples, we turn to our next research question:

\RQ{Are associated edits important for code update prediction, or 
simply serve as few-shot examples for the code update task?}

In other words, is there something to be gained by using associated edits beyond what
we gain by just adding some few-shot examples of code updates (that are not necessarily associated)?

\begin{table}[t]
  \caption{Less relevant edits degrade prediction:
  \davinci on filtered \cpo test set for different associated edits.}
  \label{tab:rq3}
  \vspace{-2ex}
  \small
\begin{tabular}{ccc|c}
\toprule
\multicolumn{3}{c||}{Choice of Associated Edits}  & Exact Match \\
\cline{1-3}
\multicolumn{1}{c|}{Association} & \multicolumn{1}{c|}{Dataset} & \multicolumn{1}{c||}{Repository} &  \\ 
\midrule
\multicolumn{1}{c|}{Spatial}  & \multicolumn{1}{c|}{Filtered} & \multicolumn{1}{c||}{Same} & {\bf{67.92}}  \\  
\multicolumn{1}{c|}{Random} & \multicolumn{1}{c|}{Filtered} & \multicolumn{1}{c||}{Same} & 64.90 \\ 
\multicolumn{1}{c|}{Random} & \multicolumn{1}{c|}{Filtered} & \multicolumn{1}{c||}{Other} & 55.82  \\ 
\multicolumn{1}{c|}{Random} & \multicolumn{1}{c|}{Unfiltered} & \multicolumn{1}{c||}{Same} & 43.23  \\ 
\multicolumn{1}{c|}{Random} & \multicolumn{1}{c|}{Unfiltered} & \multicolumn{1}{c||}{Other} & 43.64 \\
\midrule
\multicolumn{3}{c||}{ No Associated Edits}  & 37.09  \\ 
\bottomrule
\end{tabular}
\end{table}


\paraheader{Results}
Table~\ref{tab:rq3} shows the \emph{Exact Match} we get using the \davinci model using
different sets of edits as the ``associated edits''. We use the \cpo filtered test set again for
evaluation. We saw before that we get a $67.92\%$ Exact Match on this test set (Table~\ref{tab:rq1}). 
This case corresponds to when the prompt includes spatially close filtered
edits from the same file (this is a property of the \cpo benchmarks).
Let us now randomly sample edits to include in the prompt.
There are two dimensions and two buckets in each dimension to use for sampling: 
the filtered dataset versus the unfiltered dataset, and 
edits from the same repository versus edits from different repositories.
%
Randomly picking filtered edits from the same repo drops performance only slightly.
However, randomly picking filtered edits from other repos drops performance more significantly to $55.82\%$.
When sampling from unfiltered edits - irrespective of whether edits are from the same or different repos - the Exact Match remains consistently around $43\%$. 
We recall that when we provide no associated edits in the prompt, we had $37.09\%$ Exact Match (Table~\ref{tab:rq1}).

The results show that going from filtered to unfiltered edits reduces relevance of edits to the target {\em{filtered edit}}. This is because
the filtering step in~\cite{c3po} actually removes certain kinds of edits; for example, edits that are pure insertions or deletions, or edits that result in unparseable code. Hence, a randomly picked unfiltered edit is more likely to be structurally different from our target edit (which was picked from the filtered test set.)

The results also show that picking edits from repositories other than the repository of the target edit reduces relevance of the edit to the target edit. This is because
 edits from the same repository
could potentially be using common concepts, classes, methods, programming practices, and even contain similar changes.

Finally, we note that using unfiltered edits from other repositories ($43.64\%$) is still better than not using them
($37.09\%$).
This is possibly due to the LLM leveraging its \textit{few-shot} learning capabilities in that case. The gain from
around $43\%$ to around $68\%$ can thus be attributed to the associated edits.
We can, therefore, conclude that:

\RS{2}{Associated edits play a crucial role in predicting a target edit, and the Exact Match metric drops
as the relevance of the edits to the target edit drops.}

\subsection{Pre-trained Outperforms Custom}

When working with code and code edits, LLMs (such as \codex and \gptth) and other pre-trained models (such as \codet)
use byte-pair encodings (BPE) to 
tokenize code and then represent code as a sequence of tokens -- in the same way as Natural Language is represented.
In contrast, some works have argued for the use of custom representations for code and code edits
that partly capture the
parse structure and/or the programming language semantics. The paper that introduced the \cpo dataset~\cite{c3po}
also used the spatial edits used by our pre-trained LLMs but they learned a custom model employing code-centric representations for code edits. 
Our next research question
concerns comparing our approach based on pre-trained models with prior work on custom neural approaches. While both the approaches have access to the associated spatial edits, we want to understand how pre-trained models with their text-based prompts compare against models with custom code representations.


\RQ{How does our LLM-based approach compare with the \cpo approach based on a custom neural model on
the associated code update task?}


\begin{table}[t]
  \caption{Comparison with \cpo.}
  \label{tab:rq2}
  \vspace{-2ex}
  \small
  \begin{tabular}{l||c|c}
    \toprule
    &  \multicolumn{2}{c}{Exact Match on} \\
    Model &  \cpo & \overwatch \\
    \midrule
    \cpo & {{53.20}} & 10.50 \\
    \davinci & {{67.92}} & \textbf{49.09} \\
    \codetu & {{74.16}} & 34.00\\
    \codetuf / \codetuo & {\bf{81.83}} & 48.23\\
    \bottomrule
  \end{tabular}
  \vspace{0.5ex}
\end{table}

Let us compare how the \cpo custom neural model performs in comparison to \davinci 
and fine-tuned \codet. 
We first compare these models on the filtered \cpo test set and the \overwatch test set. 
Table~\ref{tab:rq2} shows that both \davinci and fine-tuned \codet 
significantly outperform the custom \cpo model on both test datasets.
The \cpo model was reported to give a 53.2\% accuracy~\cite{c3po} on the \cpo test set,
whereas both \davinci and fine-tuned \codet give better results. 
The \codetuf model gives $81.83\%$ Exact Match, which is
significantly higher than $53.2\%$ achieved by the $\cpo$ model.
Similarly, on the \overwatch dataset, the best possible configuration of \cpo was reported to give
$10.5\%$ Exact Match~\cite{overwatch}, whereas all of \davinci ($49.09\%$), \codetu ($34\%$),
and \codetuo ($48.23\%$) perform significantly better.

\ignore{
\begin{table}[t]
  \caption{\codetu performance on C3PO unfiltered test set.}
  \label{tab:c3po-unfiltered-results}
  \begin{tabular}{l||c}
    \toprule
    Experiment & Exact Match \\
    \midrule
    Without Associated Edits &   45.30 \\
    \midrule
    With Associated Edits  & \textbf{57.30} \\
    \bottomrule
  \end{tabular}
\end{table}
}

\paraheader{Comparison on unfiltered \cpo test set}
The \cpo model does not report results on the unfiltered \cpo dataset. This is partly because it contains benchmarks that are out of scope for their technique. Two such notable benchmarks are: (a) benchmarks that contain {\em{alien insertions}} where the inserted code contains tokens that do not occur in either the associated edits or the current version of the target code snippet, and (b) benchmarks that contain code snippets that cannot be parsed by an underlying parser (this step is important for \cpo to generate the Abstract Syntax Tree (AST)).
\grace can handle both these classes of benchmarks.
We evaluated \davinci on
a 5.9K random sample from this test set 
and obtained a 43.47\% Exact Match. These 5.9K samples did not contain any 
benchmarks from the filtered set. (We used a sample because of the cost of doing inferences using an LLM.)
On the {\em{full}} unfiltered \cpo test set, \codetu has $57.3\%$ Exact Match using
\grace and $45.30\%$ without.
These numbers are lower than those for the filtered \cpo test set. This indicates
that the unfiltered benchmarks are more challenging than the filtered benchmarks, which is
at odds with the informal assertions to the contrary in~\cite{c3po}.

\paraheader{Alien Insertion Benchmarks}
We extracted the samples from unfiltered \cpo test set that involved alien insertions. On that set, 
\codetu with 
\grace
achieved 17.6\% exact match, but only 10.28\% without it. The \davinci model achieved 17.37\% exact match with 
\grace
and 10.67\% without it. This indicates that 
conditioning on associated edits can help with hard benchmarks.

Admittedly, the \cpo model is much smaller (750K parameters) compared to
both \davinci (175B) and \codet (220M). However, these large models are
pre-trained and hence they can be quickly fine-tuned or prompt engineered for downstream tasks without 
need for excessive training data. Furthermore, the pre-trained models are not limited in scope, as we have discussed above.


\RS{3}{Pre-trained language models can be tuned to yield higher Exact Match compared to 
 the custom \cpo model for associated code update prediction.}

\subsection{Temporal Edit Prediction}
\label{subseq:overwatch}

\begin{table}[t]
  \caption{Comparison with \overwatch.}
  \label{tab:overwatch-apps-results}
  \vspace{-2ex}
  \small
  \begin{tabular}{l||r|r|r|r}
    \toprule
    Technique  & \overwatch & \tabular{c} \textsc{Codex-}\\ \textsc{Davinci} \endtabular & \codetu & \codetuo \\
    \midrule
    Exact Match & 38.17 & \textbf{49.09} & 34.00 & 48.23 \\
    \bottomrule
  \end{tabular}
\end{table}


The temporal edit prediction problem is an application that is well-suited for using 
\grace.
%
The state of the art in this application domain is \overwatch~\cite{overwatch}.
Fundamentally, \overwatch is solving a different problem from the associated edits prediction
problem: the input to \overwatch are \emph{fine-grained IDE version
histories} of the form $v_0, \ldots, v_n$ where each $v_i$ is a version of the
source code file.
The edit histories are extremely fine-grained, at the keystroke unlike source control
histories.
For example, if a developer types a variable name \texttt{predicate}, each 
intermediate file version containing the prefixes \texttt{p}, \texttt{pr}, \ldots is
present in the edit history.

The \overwatch technique takes the set of such IDE version histories as a training set, and
produces a ranked sequence of edit sequence patterns (ESPs).
At inference or run-time in an IDE, each ESP examines the current version history
$v_0, \ldots, v_n$ and
(a)~identifies a sequence of transitive coarse-grained edits
  $\delta_{i_0,i_1}, \delta_{i_1,i_2}, \ldots, \delta_{i_{k-1},i_k}$ 
  (i.e., each $\delta_{i_0,i_k}$ is the edit between the potentially non-consecutive
  versions $v_{i_j}$ and $v_{i_{j+1}}$), and 
(b)~uses these edits to predict the next edit to $v_{i_k}$.
%
In short, the ESPs are doing two tasks: (a) identifying ``associated edits'' from
fine-grained version histories, and (b) using these associated edits to predict the next
edit.
\RQ{Can our LLM based approach be used in conjunction with \overwatch's temporal associated edit identification? How does it compare with \overwatch's symbolic edit prediction component?}
The second task above is exactly the prediction from associated edits problem we are
tackling in this paper.
Hence, we run \overwatch on its test data of $399$ version histories with over $200,000$
versions, and gather the associated edits wherever the ESPs are able to identify them.
This results in a dataset of 1048 cases as mentioned in Section~\ref{sec:setup}.
At training time, \overwatch identifies a set of 9.9K edit sequences from the older data of $682$ version histories, however using different techniques.
The edit sequences are such that each of them belong to some commonly occurring edit sequence pattern across version histories -- they are the supports for ESPs -- and thus each of them can be treated as a set of associated edits (all edits in the sequence but the last), and
expected edit prediction (last edit in the sequence).
We use this set of 9.9K instances to further fine-tune \codetu to obtain \codetuo; see Table~\ref{tab:models}.

%
Table~\ref{tab:overwatch-apps-results} summarizes the different models' performance on the $1048$ test cases, along with \overwatch's
predictions as a baseline.
Except \codetu, 
all of our models beat the prediction component of \overwatch by a considerable margin of roughly 10\%.
\RS{4}{Our LLM-based techniques, in conjunction with systems like \overwatch creates neuro-symbolic solutions that are better at predicting next edit compared to purely symbolic techniques.}

\vspace{-2ex}
\subsection{Qualitative Analysis}
\label{sec:discussion}
Our experiments support two major observations:
(a) LLMs can predict edits that existing techniques fundamentally cannot support, and
(b) the addition of associated edits improves the performance of LLMs on the task of predicting code edits.
Next, we provide insights into why these observations hold true.

\subsubsection{\large Comparison with existing techniques}
In the following few paragraphs, we discuss the salient features of LLMs and 
the \grace
prompt design that help our approach outperform existing techniques, i.e., \cpo and \overwatch on certain kinds of edits.

\paraheader{ Generative capabilities of LLMs are useful in predicting alien insertions}
As discussed in Section~2, the LLMs we discuss in this paper can support most forms of insertions as they have access to a wide number of tokens through their pre-training and our prompting setup doesn’t restrict the tokens that the models can generate. Existing techniques are restricted in this aspect by design: {\em{\cpo cannot insert tokens other than those found in the contextual edits and Overwatch may learn patterns where the prediction template is incomplete due to unavailable mappings for holes in the Temporal Edit Pattern.}} 
For instance, consider the scenario in Figure~\ref{fig:errorcode} where a developer is trying to add HTTP error codes to error reporting calls. Here, the developer first edits Line~3 by adding a \linecode{BadRequest} error code to the reporting call. They then move to Line~5 to make a similar edit. Note that the expected error code on Line~5 is different from the one on Line~3 as it corresponds to a “No such action” error message. Moreover, the expected error code has a token ‘NotFound’ which is not present anywhere in the existing context. As C3PO’s pointer network can only pick paths to/from existing nodes, it cannot generate this new token. \davinci with \grace can correctly predict this edit.


\begin{figure}
\lstset{
  firstnumber=1,
  basicstyle=\scriptsize\ttfamily,
  morecomment=[l][\color{darkgreen}]{+\ },
  morecomment=[l][\color{red}]{-\ }
}
\begin{lstlisting}
catch (Exception ex) 
{
-  info.ReportClientError('Scheme is missing');
+  info.ReportClientError('Scheme is missing',System.Net.HttpStatusCode.BadRequest);
}
default: 
-  info.ReportClientError('No such action');
+  info.ReportClientError('No such action',System.Net.HttpStatusCode.NotFound);
\end{lstlisting}
\vspace{-2ex}
\caption{User adds \linecode{BadRequest} error code on Line~3 and moves to Line~5 where we should predict inserting \linecode{NotFound}.}
\label{fig:errorcode}
\vspace{2ex}
\end{figure}

\paraheader{Access to local spatial context in the prompt is useful}
\overwatch learns patterns and templates from observed edit sequences and strictly relies on these patterns to make predictions. There are cases, however, where the pattern learnt by \overwatch is too general and is applicable irrespective of what is in the spatial vicinity of the target edit. 
To better understand this limitation, consider the scenario in Figure~\ref{fig:ex.Input} from an active IDE editing session. The user first replaces \linecode{ex} on Line~5 with \linecode{ex.Output} and then moves to Line~3 to make the next edit. 
%
%
\begin{figure}
\lstset{
  firstnumber=1,
  basicstyle=\scriptsize\ttfamily,
  morecomment=[l][\color{darkgreen}]{+\ },
  morecomment=[l][\color{red}]{-\ }
}
\begin{lstlisting}
foreach (var ex in currentExamples)
{
-  Console.WriteLine(GetText(ex, diff.BeforeFile));
+  Console.WriteLine(GetText(ex.Input, diff.BeforeFile));
-  Console.WriteLine(GetText(ex, diff.AfterFile));
+  Console.WriteLine(GetText(ex.Output, diff.AfterFile));
}
var output = Run(currentExamples.First().Input);
AssertEqual(currentExamples.First().Output, output);
\end{lstlisting} 
\vspace{-2ex}
\caption{{User replaces \linecode{ex} on Line~5 by \linecode{ex.Output}, moves to Line~3 where we should predict replacing \linecode{ex} by \linecode{ex.Input}.}}
\vspace{1ex}
\label{fig:ex.Input}
\end{figure}
%
Overwatch gets triggered on Line~3 and predicts that \linecode{ex} should be replaced by \linecode{ex.Output} since it learns the pattern ``repeat the same replacement'', which is an instance of a common pattern. 
However, this is incorrect since \linecode{ex} should be replaced by \linecode{ex.Input} here. \davinci, with 
\grace
, correctly predicts this edit because \linecode{ex} on Line~3 is followed by \linecode{diff.BeforeFile} and Line~8 has additional information about the property \linecode{Input} that is associated with each entry in `currentExamples'. Our prompt design allows flexible addition of this additional spatial context through the {\linecode{<Prefix>}} and {\linecode{<Suffix>}} tags. \overwatch, on the other hand, cannot access spatial context that is not already present in the learned template.

\paraheader{Language-based pre-training is useful in identifying semantic editing patterns}
Scenarios in Figures~\ref{fig:errorcode} and~\ref{fig:ex.Input} also highlight the ability of LLMs to predict patterns based on semantics of identifiers in the context. In Figure~\ref{fig:ex.Input}, \davinci seems to understand that the relationship between \linecode{diff.AfterFile} and \linecode{diff.BeforeFile} would also reflect in the preceding argument (\linecode{ex.Output} and \linecode{ex.Input}, respectively). In Figure~\ref{fig:errorcode}, \davinci uses the signal from the \linecode{'No such action'} error message to correctly predict that the error code should be \linecode{System.Net.HttpStatusCode.NotFound}. Existing techniques such as \cpo and \overwatch rely on edit path analogies and symbolic editing patterns respectively to understand the editing intent. Without the use of a language-based pre-training component, it may be difficult to obtain the semantic understanding needed to perform the edits in Figures~\ref{fig:errorcode} and~\ref{fig:ex.Input}.

\vspace{-1ex}
\subsubsection{\large Benefits of using associated edits}
We observed three key benefits of providing associated edits to LLMs:


\paraheader{Associated edits help in clarifying the editing intent of the developer}
The illustrative example in Section~\ref{sec:motivation} (Figure~\ref{fig:ex_0}) showed that associated edits provide strong signals about the next edit that the developer intends to perform. In fact, without associated edits, \davinci doesn't predict the right edit even in the top-5 results. With associated edits, the correct prediction is ranked at the top suggesting that associated edits help improve the top-1 performance of the model. 

\paraheader{Associated edits emphasize relevant code context}
While LLMs like \davinci can support a large number of tokens in their prompts (~4K in \davinci's case), it has been observed that irrelevant information in the prompt affects the model's ability to attend to the right set of tokens \cite{Shi2023LargeLM}. In the illustrative example in Figure~\ref{fig:ex_0}, the target edit is 247 lines away from the required spatial context. \davinci can predict the right import with only 4-5 lines in the spatial context and access to the associated edit. The scenario without associated edits, on the other hand, requires providing ~250 lines of mostly irrelevant code to the model to include the Exception that the required import provides. \davinci fails to generate the right prediction in the top-5 results even with all of this spatial context. On a simpler version of this example where the relevant code context is moved closer to the target edit (from Line~250 to Line~15), \davinci without associated edits predicts the right import in top-10, but it is not the top-1 prediction.

\paraheader{Associated edits contain information about edited code elements}
There may be key variables that are deleted or replaced by previous edits but referenced by the target code location. Without access to these associated edits, the model has no context about these variables, methods or other code elements. For example, if a variable \linecode{var1} is replaced by \linecode{var2} in a previous edit and the developer now moves to line \linecode{var1 = var1 / 2}, the model is expected to replace this line with \linecode{var2 = var2 / 2}. Without access to the previous edit, the model doesn't know the relationship between \linecode{var1} and \linecode{var2} and may consider them to be two distinct variables.

\vspace{-2ex}
\subsection{Additional Results \& Discussion}
We conducted additional experiments to understand how 
\grace
affects \textit{robustness} and \textit{entropy} during prediction.
We also evaluated other prompting styles and model configurations.
See the technical report for details and further discussion.

\ignore{
\todo{Some observations that we may or may not want to discuss in the paper.}

We have four categories of scenarios based on whether we failed or succeeded to 
predict with or without the associated edits.

\paragraph{Succeed with associated edits, but not without}
We performed manual inspection of some randomly selected instances where the use of 
associated edits was critical to getting the correct prediction.
The general observation was that in these benchmarks, the model with associated
edits is able to learn a program transformation rule (even from noisy demonstrations)
that it is able to successfully apply on the target code.
Most typical examples of such program transformations are variable
renamings, type changes, and method signature changes, but we also 
see examples of control flow changes (removing or addition of a if or using block)
and logic changes (conditional expression transformations).

\begin{itemize}
\item
The misaligned benchmarks.
\item
The cut-paste scenario.
\item
Remove the usage of a variable followed by removing it from the arg of a function.
\end{itemize}

\paragraph{Succeed without associated edits, but not with}
If we consider those cases where the model predicted correctly without edit context, but
predicted incorrectly with edit context, then in
most such cases, the prediction with edit context was reasonably close to the ground truth,
even though it did not match it exactly.

In fact, in some cases the ground truth did exist in the top-3 predictions, but it was
not the top one obtained when using an edit context.
We also noticed a few cases where there was extraneous code in the prediction before or after the 
ground truth snippet. There were also a few cases where the prediction with edit context
was correct also included a gratuitous name change that was not required. In many such cases,
the new names were meaningful but they were not the ones the developer had used.

There were only few cases where the previous state misleads and results in an incorrect
prediction. For example, in one case the previous version contained expressions creating
classes such as ``SelectExpression'' which lead the model to use these types instead of
the more generic ``Expression'' type that was present in the ground truth. 
(EntityFrameworkCore.74827.1.)
Another challenging case was when the contextual edit involved both removing the use of a 
method and also adding an application of another method with the same base name
(ServiceStack.31816.1).
Sometimes there was an edit in the edit context that was very similar to
the one required to predict the ground truth, but it resulted in an incorrect prediction.
(EntityFrameworkCore.93218.1.)

\paragraph{Fail both with and without associated edits}

One class of examples where the model fails to predict the ground truth -- irrespective
of whether is is provided associated edits in the prompt -- are where we need to 
generate entirely new code or a new name. For example, 
there are examples where we need to add ": ILanguageBasedService" in the code,
but the model guesses the wrong names since the correct one does not appear anywhere in
the context.

Another such example is where we set a new variable ``isMainEntry = ...'' and then we 
need to predict the change on ``command.AddEntry(entry)".
Our model predicts ``if (isMainEntry) command.AddEntry(entry)" which is one reasonable guess.
Another guess would be the ground truth ``command.AddEntry(entry, isMainEntry)". 
There is nothing in the context to infer the signature of ``AddEntry" so model is unable to pick the
correct prediction. Adding a small usage example of "command.AddEntry" with 2 arguments might
help here. This shows that even associated edits may not be enough, and we may need relevant code context 
too.
\endignore}

\vspace{-1ex}
\section{Related Work}
\label{sec:related}

\subsection{Automatic code editing} 
In recent years, there has been a significant boom in academic and industrial research for automating developers' code editing activities. Most modern IDEs~\cite{visualstudio, eclipse} support automated code changes like the addition of boilerplate code, developer-assisted refactoring, etc. While these developer-assisted approaches tremendously help boost productivity~\cite{moser2008case}, a significant amount of further research exists in automated code editing aimed at learning code edit patterns from developer's previous edits ~\cite{nguyen2014statistical, nguyen2016api, tansey2008annotation, andersen2012semantic, raychev2013refactoring, ge2012reconciling, Chakraborty2018CODITCE, sequencer, codeitT5, c3po, overwatch, editpro}. We divide these approaches into two orthogonal directions:

\paraheader{Symbolic approaches}
Symbolic approaches learn the code transformation patterns by representing the example edits with symbolic abstractions. 
Given a set of such symbolically represented abstract edits, these approaches generalize the edit patterns as a sequence of edit operations. 
For instance, Refazer~\cite{refazer} represents syntactic changes with Domain Specific language and uses a deductive inference algorithm to generalize and synthesize common edit patterns. 
More recently, Overwatch~\cite{overwatch} learns to generalize developer code editing behavior from a sequence of code versions. 
Each edit is represented as {\em pre} and {\em post} program states, and generalized edit sequences are derived from an edit graph from these state pairs. 
While the earlier works in symbolic editing~\cite{meng2013lase, refazer, meng2011sydit, witch_doctor, resharper} primarily focused on syntactic editing, i.e., refactoring, similar to \overwatch~\cite{overwatch}, we also focus on semantic changes in code. Similar to \overwatch, we emphasize on conditioning future edit \wrt associated edits. 
However, unlike \overwatch, \grace does not necessarily need demonstrations of the specific edit sequence pattern to learn to apply that pattern.

\paraheader{Neural Network-based approaches}
Recent advancements in machine learning and neural networks have catapulted the field of code editing with Neural Networks relying on their noise tolerance and generalization capabilities. 
As such, several approaches~\cite{c3po, Chakraborty2018CODITCE, codeitT5, sequencer, tufano2019learning, dinella2020hoppity} have been proposed over the years using different types of Neural Networks for automatically generating edits. 
Notable among these are Sequence to Sequence Neural Machine Translation based approaches~\cite{tufano2019learning, sequencer, chakraborty2021multi}, Tree to Tree translations approach~\cite{Chakraborty2018CODITCE}, and Graph Neural Network based approach~\cite{dinella2020hoppity, yin2019learning}.
While most of these approaches learn to generalize code edit patterns from seemingly unrelated example edits, this work shows the importance of related associated edits. 
Nevertheless, the most notable feature of Neural Code Editing approaches is how the approach generates the edited code. 
While some approaches~\cite{dinella2020hoppity, c3po, codeitT5} generate a script of edit operations (\ie insert, delete, update), others~\cite{tufano2019learning, sequencer, Chakraborty2018CODITCE} generate the edited code applying the edit pattern in the process of translation. 
Similar to the latter approach, we generate the edited code given the code before the edit. 

\vspace{-2ex}
\subsection{Deep Learning for source code}
Recent advancement in Deep Neural Networks (DNN) has drawn focus on the application of such in different source code understanding and generation tasks, including bug detection~\cite{chakraborty2021deep, ding2022towards}, code comprehension~\cite{ahmad2020transformer}, code search~\cite{cambronero2019deep}, code generation~\cite{codet5}, code translation~\cite{lachaux2020unsupervised, ahmad2022summarize}, program repair~\cite{tufano2019learning, sequencer}, etc. 
The vast plethora of DNN models in SE tasks ranges from general-purpose models~\cite{ahmad2021unified} inspired by Natural Language Processing to custom-built models for source code modeling~\cite{guo2020graphcodebert, dinella2020hoppity, yin2019learning}. 
These models, however, require a large quantity of labeled data to optimize millions of parameters.
To overcome this problem, researchers have proposed to {\em pre-train} models with a large quantity of unlabelled data and subsequently re-use such a pre-trained model across different tasks~\cite{devlin2018bert, radford2018improving}. 
There are a wide variety of pre-trained models for source code proposed over the years~\cite{feng2020codebert, ahmad2021unified, codet5, codex}, some containing hundreds of billions of parameters~\cite{codex}, colloquially known as large language models  or LLMs. 
LLMs show excellent promise in autonomously learning programming language properties and additionally, have shown the ability to learn deductive reasoning inherent in programming and natural languages~\cite{wei2022chain, manning2022human, zelikman2022star, rytting2021leveraging}.
As such, these LLMs are leveraged in many industrial developer assistance tools such as GitHub Copilot~\cite{copilot}, Amazon CodeWhisperer~\cite{amazon_codewhisperer}, Intellicode Compose~\cite{intellicode_codmpose_software, intellicode}, etc. 
In this work, we show an in-depth investigation of harnessing the power of these LLMs for automated code editing.

\vspace{-1ex}
\section{Limitations \& Threats to Validity}
\label{sec:threats}
\paraheader{Limitations}
There are certain limitations of 
our approach
that we would like to address in future work. Firstly, as our approach depends on other edit mining techniques, it is restricted by the quality of the collected edits. On rare occasions, associated edits in the prompt can also mislead the model with some irrelevant information which in turn leads to incorrect predictions. Moreover, our approach can also fail when the ground truth requires knowledge of certain context (method signatures, for example) that does not appear in the associated edits. Secondly, the LLMs used are prone to known issues such as hallucinations, generation of uncompilable code, etc. Despite being generative, these models can still fail to predict edits that involve generating entirely new code. 

\paraheader{Threats to validity}
When using a pre-trained model, there is always a threat of test data leaking to the train set \cite{code-duplication}.
It is possible that the data used for pre-training \davinci contained some or all of the data in the \cpo test set since the \cpo dataset was created from GitHub repositories. One way to mitigate this threat is to perform evaluation on multiple test sets. Therefore, we also performed our evaluation on the \overwatch dataset. The \overwatch test set  was not publicly available and we obtained it directly from the authors. Hence, we believe our results are not inflated because of the possibility of \davinci having seen the \cpo test set. We mitigated the threat further by performing the same experiments on fine-tuned \codet. All conclusions we make in this work are informed by results from both models on both datasets. Finally, this potential data leak would affect all our experiment settings with the \davinci model equally and any benefit would also have been available to the model without associated edits. Our results suggest that the model clearly benefits from the addition of associated edits thus entailing a fair comparison.  

The test sets are another source of possible gap between what we observe in our experiments and what we may see if the approach were deployed in real world.  The \cpo dataset was created from commits.  It defined an edit at a certain level of granularity. This definition may not match the notion of edits used in some target application (of our code prediction models). Again, we mitigate this threat by also testing on \overwatch dataset that uses a different level of granularity for defining an edit. Our results appear to hold across the different possible notions of an ``edit''.
In fact, by presenting the associated edits to the model (in the prompt and during fine-tuning), we are able to teach the notion of an edit to it. Even with the notion of edit conveyed, the distribution of associated edits in our test sets may not reflect what we observe in practice. The approach based on \davinci is not immune to this threat, but the fine-tuning approach can adapt if we have fine-tuning data.
\vspace{-5ex}
\section{Data Availability}
\label{sec:data_availability}
The \cpo dataset is made publicly available by the authors of~\cite{c3po}. We share the scripts, prompts, and instructions to access the fine-tuned models on \cpo at \hyperlink{https://aka.ms/GrACE-Code}{\small\texttt{https://aka.ms/GrACE-Code}}. Since the \overwatch dataset is private, we do not hold the authority to redistribute the dataset or any models learned from that dataset. Readers with access to \overwatch data can reproduce the experiments using the shared scripts. An interactive tutorial notebook discussing deployment of our approach in an IDE-based edit suggestions tool is also available at the same webpage.

\vspace{-2ex}
\section{Conclusions \& Future Work}
\label{sec:conclusion}
Predicting code edits is an important software-engineering problem. In this paper, we leverage the 
generative capability of LLMs to address this problem. Without the knowledge of prior edits, the
LLMs fail to predict the required edits, but when we combine them with associated edits, their
performance improves greatly. This simple strategy is quite effective, and as shown in the experiments,
\grace outperforms the current state-of-the-art specialized symbolic and neural methods on their respective datasets.

The generative capability of LLMs has opened up many opportunities for addressing software-engineering
problems that have been hard to deal with. We believe that combining the LLMs with domain-specific insights,
such as our use of associated edits, holds promise for hitherto challenging problems. In the future, we shall
seek to exploit this strategy for other software engineering problems. On the problem of predicting code edits,
we plan to explore the problem of discovering associated edits, and the application to large-scale migrations, refactorings, and maintenance activities.



\bibliographystyle{ACM-Reference-Format}
\bibliography{main}

\balance
\newpage

\appendix

\section{Motivating Examples}

We present an illustrative example that has {\em{two}} associated edits which are both
used to make a final prediction using a fine-tuned CodeT5 model.


\begin{figure*}
\vspace{-2ex}
\centering
\lstset{
  firstnumber=1,
  basicstyle=\footnotesize\ttfamily,
  morecomment=[l][\color{darkgreen}]{+\ },
  morecomment=[l][\color{red}]{-\ }
}
\begin{lstlisting}
-  return AssertQueryAsync<Order>(  
+  return AssertQuery<Order>(
+    isAsync
    os = > os.Where( o = > o.OrderDate != null  &  &  o.EmployeeID.Contains( "10" ) ) 
             .Select( o = > new Order{ CustomerID = o.CustomerID } ) ,  
    e = > e.CustomerID ); 
}

-  [ ConditionalFact ] 
+  [ Theory ]
+  [ InlineData( false )]
+  [ InlineData( true )]
-  public virtual Task Select_expression_other_to_string(  )
+  public virtual Task Select_expression_other_to_string( bool isAsync )
{  
-      return AssertQueryAsync<Order>( 
+      return AssertQuery<Order>(
+        isAsync
        os = > os.Where( o = > o.OrderDate != null ) 
             . Select( o = > new Order{ ShipName = o.OrderDate.Value } ) ,  
        e = > e.ShipName ); 
} 

-  [ ConditionalFact ]
+  [ Theory ]
+  [ InlineData( false )]
    ...
\end{lstlisting} 
\vspace{-2ex}
\caption{{User replaces \linecode{AsserQueryAsync} on Line~1 with \linecode{AssertQuery} and adds an argument \linecode{isAsync} (Line~3). The next edit replaces \linecode{[ConditionalFact]} on Line~24 with \linecode{[Theory], [InlineData(false)]}. The task is to predict edits on Lines~9-18.}}
\vspace{1ex}
\label{fig:motivation1}
\end{figure*}

Consider a developer refactoring code as shown in Figure~\ref{fig:motivation1}.
The developer first performs edits on Line~1 to turn the asynchronous function
\linecode{AssertQueryAsync<Order>}
to the function
\linecode{AssertQuery<Order>} that takes one extra argument - a Boolean flag 
\linecode{isAsync} (Line~3).
The second edit performed by the developer involves replacing the attribute
\linecode{[ConditionalFact]} on Line~24 by
new attributes
\linecode{[Theory]} and
\linecode{[InlineData(false)]} on Lines~25,26.
After performing those two edits, the developer now moves to Lines~9-18.
Can we suggest the code changes to be performed on these lines and correctly predict Lines~9--18?


First, let us consider how {modern code completion tools based on LLMs} will attempt to predict the new code.
These code completion models look at the current snapshot of the code to make predictions. Note that the associated edits preserve the relationship between the \linecode{AssertQueryAsync} and \linecode{AssertQuery} and without the inclusion of these edits it would be difficult to know that the two in fact refer to the same method. Similarly, the relationship between
\linecode{[ConditionalFact]} and the new attributes \linecode{[Theory]},
\linecode{[InlineData(false)]} is also lost.
Without all that information, it will be a difficult to make the correct prediction.
However, if we have information about the two associated edits, then it is possible
to correctly predict the third edit that is needed on Lines~9-18.
In fact, given the preceding two edits, our fine-tuned CodeT5 model correctly
predicts the third edit above.
If we instead fine-tune a CodeT5 model to predict code changes just from the
code in the current version, then such a model fails to predict the correct
change.

\begin{figure}
\vspace{-2ex}
\lstset{
  firstnumber=1,
  basicstyle=\footnotesize\ttfamily,
  morecomment=[l][\color{darkgreen}]{+\ },
  morecomment=[l][\color{red}]{-\ }
}
\begin{lstlisting}
using System;
-  using NUnit.Framework;
+  using XUnit;
-  public class TimeConfigurationTests : NLogTestBase
+  public class TimeConfigurationTests : NLogTestBase, IDisposable 
{
-    [TearDown]
-    public void TearDown(){ ... }
+    public void Dispose(){ ... }
-    [Test]
+    [Fact]
  public void DefaultTimeSourceTest{ ... }
}

\end{lstlisting} 
\vspace{-2ex}
\caption{{User replaces \linecode{[Test]} on Line~10 by \linecode{[Fact]} and the \linecode{Teardown} method on Line~8 by \linecode{Dispose}. The target edit needs to replace \linecode{NUnit.Framework} on Line~2 with \linecode{XUnit} and include the \linecode{IDisposable} interface on Line~4.}}
\vspace{1ex}
\label{fig:motivation3}
\end{figure}

Figure~\ref{fig:motivation3} shows another scenario where the developer wants to migrate from the \linecode{NUnit} C\# testing framework to \linecode{XUnit}. The developer first replaces \linecode{[Test]} on Line~10 by \linecode{[Fact]} and the \linecode{Teardown} method on Line~8 by \linecode{Dispose}. Using these associated edits, our task is to predict that (1) \linecode{NUnit.Framework} on Line~2 needs to be replaced by \linecode{XUnit} and (2) the \linecode{TimeConfigurationTests} class needs to import the \linecode{IDisposable} interface on Line~4. The associated edits provide strong signals for these edits: (1) the \linecode{[Fact]} attribute is used to denote unit tests in \linecode{XUnit} (2) \linecode{XUnit} test classes need to implement a \linecode{Dispose} method for performing tasks at teardown (3) this \linecode{Dispose} method is available via the \linecode{IDisposable} interface. Our fine-tuned \codet model is able to successfully predict this change when the associated edits are provided.

Note that existing techniques such as \cpo and \overwatch may not be able to generate the right predictions on the examples in Figure~\ref{fig:motivation1} and Figure~\ref{fig:motivation3}. As \cpo can only use tokens from the context edits and the target, it will not be able to generate the new tokens required for the two edits (the scenario in Figure~\ref{fig:motivation1} needs the new \linecode{true} token on Line~12 and the scenario in Figure~\ref{fig:motivation3} needs the \linecode{XUnit} and \linecode{IDisposable} tokens on Line~3 and Line~5 respectively). Moreover, as these are not common editing patterns, \overwatch may not be able to learn templates for them. 




\section{Alternate Prompt}

The alternate comment-style prompt is shown in Figure~\ref{fig:llmprompt}.
The main difference between this prompt and the prompt described in the main
text of the paper is that we use comments in place of tags to
delineate the various parts of the prompt.

\begin{figure}[h]
\lstset{firstnumber=1}
\begin{lstlisting}
// The following piece of code is outdated.
/* <Snippet updated by 1st edit - before snapshot> */
// Here is the new version of the code.
<Snippet updated by 1st edit - after snapshot>
// The following piece of code is outdated.
/* <Snippet updated by 2nd edit - before snapshot> */
...
// Here is the new version of the code.
<Snippet updated by n-th edit - after snapshot>
// The following piece of code is outdated.
/* <Snippet at locations to be updated> */
// Here is the new version of the code.
\end{lstlisting}
\caption{Comment-style prompt for the associated code update task.}\label{fig:llmprompt}
\end{figure}

\section{Dataset statistics}

\begin{figure*}[t!]
     \centering   
    \begin{subfigure}{0.6\linewidth}
        \includegraphics[width=\linewidth]{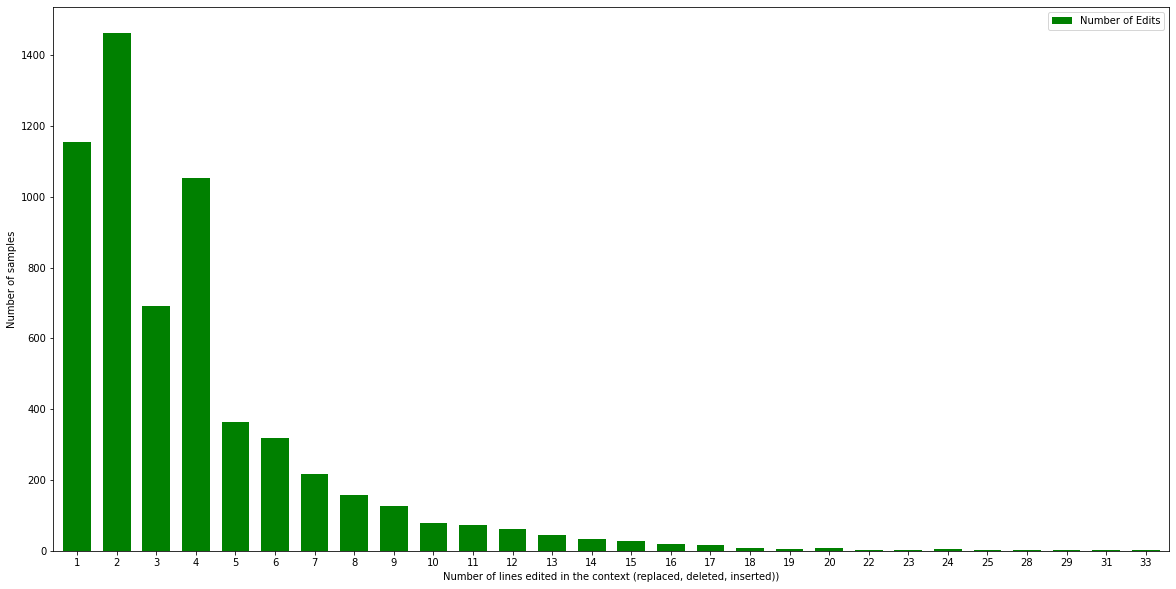}
        \caption{Distribution of number of lines edited in the context (associated edits)}
        \label{fig:edit_lengths_assoc_edits}
    \end{subfigure}
    
    \vspace{3mm} 
    
    \begin{subfigure}{.6\linewidth}        \includegraphics[width=\linewidth]{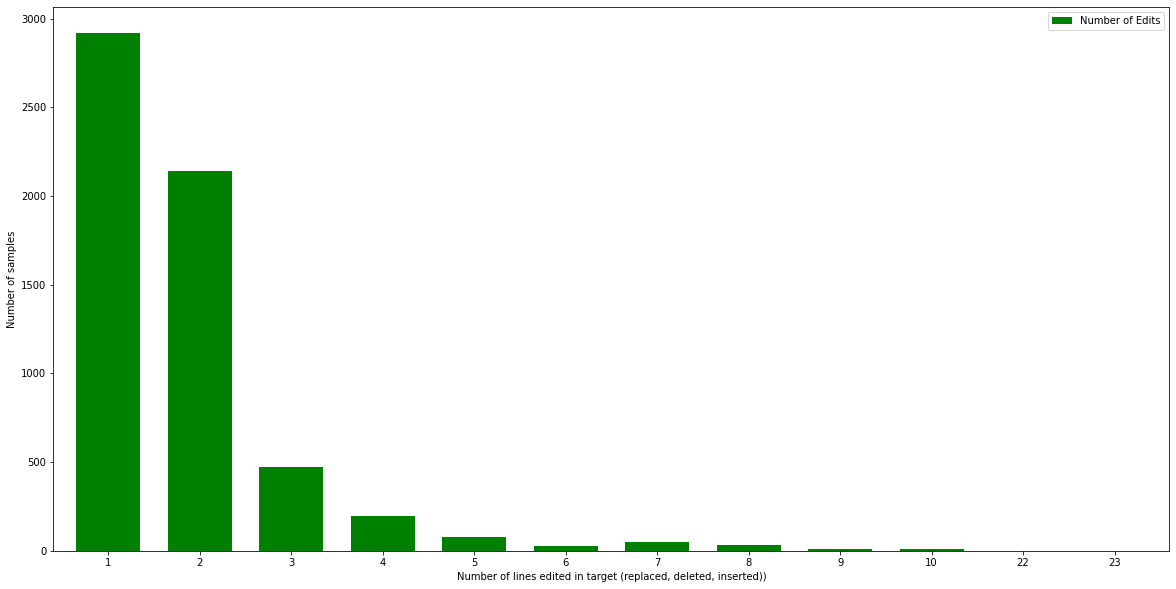}
        \caption{Distribution of number of lines edited in the target}
        \label{fig:edit_lengths_target}
    \end{subfigure}
     \caption{Edit length statistics of the \cpo filtered test set}
     \label{fig:edit_lengths}
\end{figure*}

\begin{figure*}[t!]
     \centering   
    \includegraphics[width=0.6\linewidth]{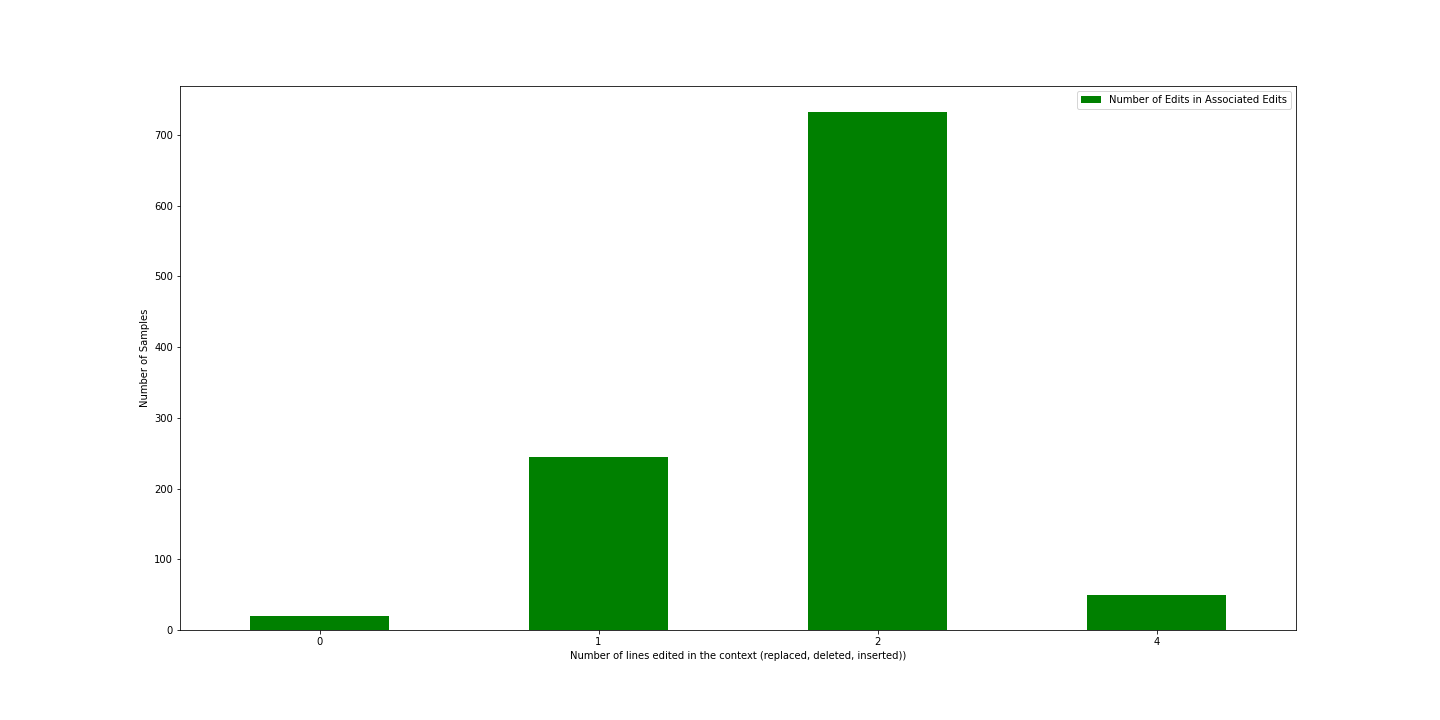}
     \caption{Edit length statistics of the \overwatch test set: Distribution of number of lines edited in the context}
     \label{fig:edit_lengths_overwatch}
\end{figure*}

There are two datasets discussed in the experiments: \cpo filtered test set and \overwatch test set. In order to discuss the difficulty of these datasets, we need to define the notion on an \emph{edit}. For simplicity, we consider an edit to be a line of code that is either replaced, inserted or deleted. Here, modification of an existing line is considered a replacement, addition of a new line is considered an insertion and removal of an existing line is considered a deletion. 

\paraheader{\cpo filtered test set}
Figure~\ref{fig:edit_lengths} shows the distribution of examples from the \cpo filtered test set in terms of the number of lines edited in the context and target. Note that the number of lines edited in the context (Figure~\ref{fig:edit_lengths_assoc_edits}) includes all the associated edits provided to make prediction. In the context, the number of lines edited on average are 3.8 and most of the samples ($90^{th} percentile$) have less than 8 line edits. The target edits are much smaller with an average of 1.8 line edits (Figure~\ref{fig:edit_lengths_target}). 90\% of the targets contain less than 4 line edits.

\paraheader{\overwatch test set}
Figure~\ref{fig:edit_lengths_overwatch} shows the edit length distribution of associated edits from the \overwatch test set. 1.8 lines have been edited in the context on average and 90\% of the samples have at most 2 edits. It is expected to see smaller edits in this dataset because it contains edits made by developers in an IDE (which are often small). Note that it is not possible to get the edit length distribution for the targets in this dataset as there is no single ground truth target edit and the predicted edit is compared with several future versions of the target (see Section~5.5 for more details on the evaluation).

\section{Additional Experimental Results}
\paraheader{Additional details about the  metrics}
In the case of \overwatch dataset, a prediction is considered an {\em{exact match}} if the {\em{version}} produced by incorporating the suggestion syntactically matches any of the 50 {\em{future versions}}.
The percentage of exact match predictions is also interchangeably called {\em{precision}}, as reported by the \overwatch paper. 
In all our results, we report {\em{Exact Match}} for Top-1 predictions and explicitly call out whenever we report Top-5 numbers (any one of the 5 is an \emph{exact match}).

Table~\ref{tab:c3po-filtered-results} presents complete
set of numbers from our various experiments
on the filtered \cpo test set that contained 5.9 benchmarks. 
Apart from the models discussed in the main text of the paper,
we also have results for a smaller Codex model \cushman.

Table~\ref{tab:overwatch-apps-test-set-results}
presents results on the 1K test benchmarks from the
\overwatch dataset.

Table~\ref{tab:overwatch-complete-test-set-results}
shows results on a edit sequence patterns test dataset that
contains 13K instances of edit sequences learned by \overwatch
on the test IDE versions.

Table~\ref{tab:c3po-unfiltered-sampled-results}
shows results obtained by \davinci\ on a random sample of 5.9K
benchmarks taken from the unfiltered \cpo test set.

Table~\ref{tab:codet5_rq3} contain data from 
\codet on filtered \cpo test set using different associated edits.
Since \codet models were fine-tuned on certain prompts, 
presenting them with prompts that look different causes a uniform
degradation of accuracy -- unlike the case for \davinci
that showed a graceful degradation as the edits
in the prompt became less relevant to the target edit.

\begin{table*}[thb]
  \caption{C3PO filtered test set (5.9K samples)}
  \label{tab:c3po-filtered-results}
  \resizebox{\linewidth}{!}{
  \begin{tabular}{cccl}
    \toprule
    \multicolumn{2}{c}{\multirow{2}{*}{Experiment}} & \multicolumn{2}{c}{Exact Match Accuracy} \\
     & & Top1 & Top5 \\
    \midrule
    & C3PO & 53 & \\
    \midrule
    Without Associated Edits & Cushman-002 (comment-style) & 3.54 & 5.59 \\
                             & Davinci-002 (comment-style) & 28.85 & 35.36 \\
                             & Davinci-002 (tag-style, temp=0.1) & 37.09 & 40.94 \\
                             & Davinci-002 (tag-style, temp=0.5) & 33.72 & 52.9 \\
                             & \codet-base (finetuned on unfiltered C3PO train) & 64.52 & 72.22 \\
                             & \codet-base (finetuned on filtered C3PO train) & 72.28 & 83.13 \\
                             & \codet-base (trained on unfiltered C3PO train, finetuned on filtered C3PO train) &  73.46 & 85.27\\
    \midrule
    With Associated Edits & Cushman-002 (comment-style) & 64.09 & 65.36 \\
    (taken from the spatial context) & Davinci-002 (comment-style) & 65.63 & 66.92 \\
                          & Davinci-002 (tag-style, temp=0.1) & 67.92 & 69.77 \\
                          & Davinci-002 (tag-style, temp=0.5) & 66.59 & 74.64 \\
                          & \codet-base (finetuned on unfiltered C3PO train) & 74.16 & 79.66 \\
                          & \codet-base (finetuned on filtered C3PO train) & 75.30 & 84.7 \\
                          & \codet-base (trained on unfiltered C3PO train, finetuned on filtered C3PO train) & 81.83 & 89.5 \\
    \midrule
    With Associated Edits & Cushman-002 & & \\
    (taken from other repos filtered) & Davinci-002 (tag-style) & 55.82 & 60.05 \\
    (taken from any repo unfiltered) & Davinci-002 (tag-style) & 43.64 & 47.58 \\
    \midrule
    With Associated Edits & Cushman-002 & & \\
    (taken from same repo filtered) & Davinci-002 (tag-style) & 64.9 & 68.09 \\
    (taken from same repo unfiltered) & Davinci-002 (tag-style) & 43.23 & 47.11 \\
    \midrule
    With Associated Edits & Cushman-002 & & \\
    (2 associated edits + 1 unrelated) & Davinci-002 (tag-style, temp=0.5) & 67.4 & 74.91 \\
                                       & Davinci-002 (tag-style, temp=0.1) & 68.51 & 70.21 \\
    \bottomrule
  \end{tabular}
  }
\end{table*}

\begin{table*}[thb]
  \caption{Overwatch applications test set (1K samples)}
  \label{tab:overwatch-apps-test-set-results}
  \resizebox{\linewidth}{!}{
  \begin{tabular}{cccl}
    \toprule
    \multicolumn{2}{c}{\multirow{2}{*}{Experiment}} & \multicolumn{2}{c}{Exact Match Accuracy}\\
     & & Top1 & Top5 \\
    \midrule
    & Overwatch & 38 & \\
    \midrule
    Without Associated Edits & Davinci-002 & 31.81 & 37.15 \\
    & \codet-base (trained on unfiltered C3PO train) & 22.25 & 48.81\\
    & \codet-base (trained on unfiltered C3PO train fine tuned on overwatch train) & 40.78 & 59.79\\
    \midrule
    With Associated Edits & Davinci-002 & 49.09 & 53.01 \\
                          & \codet-base (trained on unfiltered C3PO train) & 34.0 & 53.1 \\
    & \codet-base (trained on unfiltered C3PO train fine tuned on overwatch train) & 48.23 & 63.61\\
                          
    \bottomrule
  \end{tabular}
  }
\end{table*}
\begin{table*}[thb]
  \caption{Overwatch complete test set (13K samples)}
  \label{tab:overwatch-complete-test-set-results}
  \resizebox{\linewidth}{!}{
  \begin{tabular}{cccl}
    \toprule
    \multicolumn{2}{c}{\multirow{2}{*}{Experiment}} & \multicolumn{2}{c}{Exact Match Accuracy}\\
     & & Top1 & Top5 \\
    \midrule
    Without Associated Edits & Davinci-002 (tag-style) & 25.39 & 33.89 \\  & \codet-base (trained on unfiltered C3PO train fine tuned on overwatch train) & 17.08& 25.65\\
    \midrule
    With Associated Edits & Davinci-002 (tag-style) & 34.45 & 40.27 \\ & \codet-base (trained on unfiltered C3PO train fine tuned on overwatch train) & 21.91 & 35.21 \\
                          
    \bottomrule
  \end{tabular}
  }
\end{table*}

\begin{table*}[thb]
  \caption{C3PO unfiltered sampled test set (5.9K samples)}
  \label{tab:c3po-unfiltered-sampled-results}
  \begin{tabular}{cccl}
    \toprule
    \multicolumn{2}{c}{\multirow{2}{*}{Experiment}} & \multicolumn{2}{c}{Exact Match Accuracy} \\
     & & Top1 & Top5 \\
    \midrule
    With Associated Edits & Davinci-002 (tag-style) & 43.47 & 45.31 \\
    Without Associated Edits & Davinci-002 (tag-style) & 25.53 & 28.34 \\
    \bottomrule
  \end{tabular}
\end{table*}

\begin{table}[thb]
  \caption{Less relevant edits degrade prediction: \\
  \codet on filtered \cpo test set using different associated edits}
  \label{tab:codet5_rq3}
  \begin{tabular}{l||c}
    \toprule
    Source of associated edits & Exact Match \\
    \midrule
    Spatial & {\bf{74.16}}\\
    Any unfiltered edit from same repo & {{63.25}} \\
    Any unfiltered edit from other repo & {{63.32}} \\
    No associated edits & {{64.52}} \\
    \bottomrule
  \end{tabular}
\end{table}






\subsection{Comparison with few-shot prompting without associated edits}

In Section~5.2 of the main text, we empirically show that predictions using associated edits are significantly better than those made using randomly samples edits. These results are with the prompt design we discuss in Figure~2 (main text). While these results signify the importance of associated edits, there is another important baseline to consider: few-shot prompting without associated edits. In this case, we create a few-shot prompt using the design mentioned in Figure~3 (main text). Our preliminary analysis on the \cpo test set suggests that this approach is better than zero-shot prompting without associated edits but not as good as our approach. When \davinci is prompted with 2 random examples from the same repository (2-shot prompting), it reports an Exact Match of 45.53\%. Compared with the results in Table~5 (main text), this is expectedly better than an Exact Match of 37.07\% from the \davinci model without associated edits. It is however much lower than 67.92\% reported by \davinci with associated edits. Interestingly, this is also lower than the case where randomly sampled edits are added to the prompt in Figure~2 from the main text (64.9\%). As these two cases only differ in the way the edits are presented in the prompt, this may suggest that our prompt design actually helps the model use the provided examples better. More experiments on the unfiltered \cpo and \overwatch test sets would be needed to confirm this hypothesis.

\subsection{Associated Edits Improve Robustness}

A central thesis of our work is that code updates should be conditioned on
associated edits, and we created two models, \davinci and \codet that perform
that task.
When using those models for inference, we need associated edits. What if we don't
have any? What if we have edits that include associated edits but may have other
edits? If we use edits that are not related to the target edit, would those harm
and cause exact match to fall below if we had not used any associated edits?
If the models are very sensitive to such noise in the associated edits,
then that would increase the burden on users to find exactly the correct set of associated edits.

\RQ{How tolerant is our approach to noise in the set of associated edits provided
during inference?}

\begin{table}[thb]
  \caption{Noise tolerance of associated code update: \\
  \davinci on filtered \cpo test set}
  \label{tab:rq31}
  \begin{tabular}{l||c}
    \toprule
    Source of associated edits & Exact Match \\
    \midrule
    Spatial & {67.92}\\
    Spatial + one unfiltered edit from any repo & {68.51} \\
    \bottomrule
  \end{tabular}
\end{table}

\paraheader{Results}
We first consider the case when have one additional unrelated edit in the set of associated edits.
We added one unrelated unfiltered edit to the two spatially close associated
edits in the \cpo filtered test set. With everything else unchanged, as shown in Table~\ref{tab:rq31},
\davinci gives a $68.51\%$ Exact Match on this modified \cpo filtered test set. So, adding an unrelated
edit, in fact, slightly improved the Exact Match metric from $67.92\%$. This small gain is possibly due
to the few-shot learning feature of \davinci.

\RS{1}{Inclusion of an unrelated edit does not significantly affect the LLM's Exact Match metric on
the associated code update task.}

We next consider the following question:
How does a model trained with associated edits compare against one trained without them on the code update task? Does training with associated edits hurt the model's performance on the code update task?


\begin{table}[thb]
  \caption{Effect of associated code update finetuning on code update task: \\
  Test set: \overwatch without spatial edits}
  \label{tab:rq6-part2}
  \resizebox{\linewidth}{!}{
      \begin{tabular}{l||c|c}
        \toprule
        Technique & Finetuning Prompt & Exact Match \\
        \midrule
        \codet-base-\cpo-finetuned & without spatial edits & {22.25}\\
                                   & with spatial edits & {24.74} \\
        \midrule
        \codet-base-\overwatch-finetuned & without spatial edits & {40.78}\\
                                         & with spatial edits & {38.97} \\
        \bottomrule
      \end{tabular}
    }
\end{table}

\paraheader{Results}
In order to answer these questions, we consider evaluating two \codet models with different finetuning objectives on the \overwatch dataset. The first model is trained with prompts containing associated edits while the second one is trained without. As shown in Table~\ref{tab:rq31}, the two models report comparable results on the \overwatch dataset when no associated edits are provided. The model trained with associated edits reports a ~2.5\% higher Exact Match when finetuned on the \cpo dataset and only loses ~2\% on Exact Match when further finetuned on the \overwatch dataset. This suggests that the associated code update task can be used to train models that are robust enough to then be used out-of-the-box on the code update task. 


\subsection{Associated Edits Reduce Entropy}

Consider the probability density function (pdf) $p_1(v_n)$ on document version $v_n$
given by 
$p_1(v_n) = P(v_n | \locations, v_{n-1})$ and 
a second pdf 
$p_2(v_n) = P(v_n | \locations, v_{n-1}, \delta_{0,1},\ldots,\delta_{n-2,n-1})$.
If the associated edits $\delta_{0,1},\ldots,\delta_{n-2,n-1}$ contribute to the prediction of $v_n$,
then the entropy of $p_1$ should be higher than that of $p_2$.
The models without and with associated edits learn the functions $p_1$ and $p_2$ respectively.
The difference between Top1 Exact Match and TopK Exact Match for a given model is a good 
proxy for the entropy of the probability density function it has learnt.

Table~\ref{tab:rq5} shows that irrespective of whether we used
fine-tuned \codet, or \davinci, and independent of which training sets we
used to fine-tune \codet, when we used associated edits in the prompt,
the gap between Top1 and Top5 Exact Match was always lower than what it was
when we did not use associated edits. 
So, this indicates that associated edits are indeed predictive of code updates.

\begin{table}[thb]
  \caption{Top1-Top5 Spread}
  \label{tab:rq5}
  \resizebox{\linewidth}{!}{
      \begin{tabular}{l|l||c|c|c|c|c|c}
        \toprule
        & & \multicolumn{3}{c}{Without associated edits} & \multicolumn{3}{c}{With associated edits} \\
        Model & Test & Top1 & Top5 & $\Delta$ & Top1 & Top5 & $\Delta$\\ \midrule
        \codet 1 & \cpo & 64.5&72.2&7.7 & 74.2&79.7&{\bf{5.5}} \\
        \codet 2 & \cpo & 72.3&83.1&10.8 & 75.3&84.7&{\bf{9.4}} \\
        \codet 1 & \overwatch & {22.25} & {48.81}
    &  {24.6} & 34.0&53.1&{\bf{19.1}} \\
        \codet 12 & \overwatch & {40.78}& {59.79} & {19} & 48.2&63.6&{\bf{15.4}} \\ \midrule
        \davinci & \cpo & 37.1&40.9&3.8 & 67.9&69.8&{\bf{1.9}} \\ 
        \davinci & \overwatch & 31.8&37.2&5.4 & 49.1&53.0&{\bf{3.9}} \\
        \bottomrule
      \end{tabular}
  }
\end{table}

In Table~\ref{tab:rq5} \codet 1 is \codet fine-tuned only on \cpo unfiltered training set,
\codet 2 is \codet fine-tuned only on \cpo filtered training set,
and
\codet 12 is \codet 1 further fine-tuned on \overwatch training set.
Note that the Top1-Top5 Exact Match difference is generally smaller
for \davinci because we used a lower temperature setting (0.1) for our experiments.
%
Additional fine tuning also helps reduce entropy of the learned probability density function.
For example, the Top1-Top5 spread is extremely large for 
the case when we use \codet fine-tuned only on \cpo unfiltered training set (\codet 1)
and evaluate it on \overwatch dataset, but the spread reduces somewhat when we further fine-tune
the model on \overwatch train set (\codet 12). 
The above results provide additional evidence in support of our Result~1 (from the main text).


\subsection{Learnings from Manual Inspection}

We manually inspected a few benchmarks where some model failed to make the
correct code prediction -- either with associated edits, or without them, or in both cases.
We next share some high level observations from these investigation.

In the set of benchmarks that succeeded with associated edits, but not without,
a common case was when there was an edit in the associated edits that could be 
generalized into a program transformation rule and then applied to the edit location
to yield the ground truth.
However, often the associated edit was not a ``perfect'' demonstration of the required rule,
but only a ``noisy'' demonstration. Furthermore, there were also cases where the rule 
had to be adapted (say, by generating some new identifier based on the pattern) in
subtle ways.

In the set of benchmarks that succeeded without associated edits, but not with,
the prediction with associated edits was often very close to the ground truth.
There were a few failures (in presence of associated edits) that were caused 
due to previous versions misleading the model with some irrelevant information 
(such as, something about \linecode{SelectExpression} objects) that was close
to something in the ground truth (such as, the \linecode{Expression} type).

In the set of benchmarks that failed both with and without associated edits,
the most common case was when the edit involved generating entirely new code or a
new identifier name. In other cases, generating the ground truth would have required
knowing the signature of certain methods that did not appear in the associated edits.
It was clear that while associated edits is very useful, it is not always enough
and needs to be enhanced with ``associated code context'' (such as, live variables 
and type signatures). We leave that investigation to future work.



    




\end{document}